\documentclass[a4paper,12pt]{article}

\renewcommand{\title}[1]{
 \begin{center}
  \Large \bf #1
 \end{center}
}

\renewcommand{\author}[3]{
 \begin{center} #1 \\
  {\it #2} \\
  {\small E-mail: \texttt{#3}}
 \end{center}
}

\makeatletter
\long\def\@caption#1[#2]#3{\par
  \begingroup
    \def\\{\protect\\}
    \addcontentsline{\csname ext@#1\endcsname}{#1}%
      {\protect\numberline{\csname the#1\endcsname}{\ignorespaces #2}}%
  \endgroup
  \begingroup \@parboxrestore \if@minipage \@setminipage \fi
    \normalsize
    \@makecaption{\csname fnum@#1\endcsname}{\ignorespaces #3}\par
  \endgroup}
\makeatother
\usepackage{amssymb}
\usepackage{amsmath}

\usepackage{theorem}
\newtheorem{theorem}{Theorem}
\newtheorem{corollary}{Corollary}
\newtheorem{lemma}{Lemma}

\def\tr{{\hbox{\rm Tr}}}

\newcommand{\Slash}[1]{\ooalign{\hfil$\slash$\hfil\crcr$#1$}}

\usepackage[dvips]{graphicx}

\begin{document}

\begin{titlepage}

\baselineskip 5mm

\begin{flushleft}
February 2006
\end{flushleft}

\begin{flushright}
OCHA-PP-256 \ \ \\
hep-th/0511085
\end{flushright}

\title{Dimensional Reduction of Seiberg-Witten Monopole Equations,
 ${\cal N}=2$ Noncommutative Supersymmetric Field Theories and Young Diagrams}

\author{Akifumi Sako${}^{\dagger}$ \ , 
\ Toshiya Suzuki${}^{\S} {}^{\ast}$  \\ \ }
{${}^{\dagger}$ Department of Mathematics, Faculty of Science and
 Technology, Keio University\\
3-14-1 Hiyoshi, Kohoku-ku, Yokohama 223-8522, Japan\\ \ \\
${}^{\S}$ Department of Physics, Faculty of Engineering, Musashi
 Institute of Technology\\
1-28-1 Tamazutsumi, Setagaya-ku, Tokyo 158-8557, Japan\\ 
${}^{\ast}$ Department of Physics, Faculty of Science, Ochanomizu University\\
2-1-1 Otsuka, Bunkyo-ku, Tokyo 112-8610, Japan\\ \  }
{${}^{\dagger}$ sako@math.keio.ac.jp\\
\makebox{}\hspace{18mm} ${}^{\ast}$ tsuzuki@phys.ocha.ac.jp}


\abstract{
We investigate the Seiberg-Witten monopole equations on noncommutative
(N.C.)
 ${\mathbb R}^4$ at the large N.C. parameter limit, 
in terms of the equivariant cohomology.
In other words, ${\cal N}=2$ supersymmetric U(1) gauge
theories with hypermultiplet on N.C.${\mathbb R}^4$ are studied. 
It is known that 
after topological twisting partition functions of ${\cal N}>1$
 supersymmetric theories on N.C. ${\mathbb R}^{2D}$ are invariant under N.C.parameter
shift, then the partition functions can be calculated 
by its dimensional reduction.
At the large N.C. parameter limit, 
the Seiberg-Witten monopole equations are reduced to ADHM equations with
 the Dirac equation reduced to $0$ dimension.
The equations are equivalent to 
the dimensional reduction of non-Abelian $U(N)$ Seiberg-Witten monopole
equations in $N \rightarrow \infty$.
The solutions of the equations are also interpreted as a configuration
of 
brane anti-brane system. 
The theory has global symmetries under torus actions
originated in space rotations and gauge symmetries.
We investigate the Seiberg-Witten monopole equations reduced to
 $0$ dimension  
and the fixed point equations of the torus actions.
We show that the Dirac equation reduced to $0$ dimension is trivial when
 the fixed point equations and the ADHM equations are satisfied.
%
For finite $N$, it is known that the fixed points of the ADHM data are isolated and are
 classified by the
 Young diagrams.
We give a new proof of this statement 
by solving the ADHM equations and the fixed point equations
concretely and by giving graphical interpretations of the field components
and these equations.
}

\end{titlepage}

%
%
\section{Introduction} \label{intro}
The Seiberg-Witten theory causes a revolution of nonperturbative analysis
for ${\cal N} = 2$ supersymmetric Yang-Mills theories
\cite{Seiberg-Witten1,Seiberg-Witten2}.
In the Seiberg-Witten theory, the instanton effects of ${\cal N} = 2$ supersymmetric Yang-Mills
theories are encoded in the pre-potential,
which is defined by using the Seiberg-Witten curve.
(See, for example, \cite{some_review} and references there in.)
The Seiberg-Witten theory also provides a powerful tool, the monopole equation, to investigate
the topology of $4$ dimensional manifolds
\cite{Witten1,Witten2}.
The monopole equations are more tractable than the instanton equation, and
yield many results in  mathematics as well as physics. 

Meanwhile, instanton calculus has developed by using
ADHM data or D-instanton. (See, for example, \cite{Dorey}.)
Particularly, an important calculation technology for
${\cal N}= 2$ supersymmetric Yang-Mills theories 
is brought by Nekrasov \cite{Nekrasov}. 
After \cite{Nekrasov}, many related works have been made \cite{Fl-Po}-\cite{Fu-Mo-Po-Ta}.
In \cite{Nekrasov} and so on,
the localization theorem plays an essential role
\cite{Lo-Ne-Sh}-\cite{Mo-Ne-Sh1}. (See also \cite{DH,AB}.)
The localization theorem is valid when the
theory has symmetries which correspond to some group action 
and the group
action has isolated fixed points.
It is expected that many kinds of calculations of ${\cal N} >1$ supersymmetric gauge theory
are carried out by using this theorem.

It is shown that
partition functions of
${\cal N} >1$ supersymmetric gauge theories on
noncommutative (N.C.) ${\mathbb R}^{2D}$ are invariant under the N.C.
parameter change \cite{sako-suzuki}.
Therefore we can perform the calculation at the large N.C. 
parameter limit.
As discussed in \cite{sako-suzuki}-\cite{sako2}, taking this limit causes dimensional reduction,
and 
we can calculate the partition functions by using the theory
after dimensional reduction.
For this reason, it is important to investigate the dimensional reduction.

In this article, we will study a $0$ dimensional model given by
dimensional reduction of
Seiberg-Witten monopole equations
derived from ${\cal N}= 2$ supersymmetric $U(1)$
theory on N.C. ${\mathbb R}^{4}$.
The equations are equivalent to the ADHM equations and the Dirac equation reduced to $0$ dimension. 
The equations are also equivalent to 
the dimensional reduction of non-Abelian $U(N)$ Seiberg-Witten monopole
equations on commutative ${\mathbb R}^4$ at the large $N$ limit.
In this paper, we investigate both cases of finite $N$ and infinite $N$.
The finite $N$ case is not only the toy model,
but also the model that is possible to be implanted into
the $N= \infty$ theory and the results are valid for some special cases
of $N= \infty$ model.
We will find that
the solutions of the equations are also interpreted as a configuration
of 
brane anti-brane system. 
The theory has global symmetries under torus actions
originated in space rotations and gauge symmetries.
The torus actions define their fixed point equations.
We will investigate the fixed point equations and 
the dimensional reduction of the Seiberg-Witten monopole equations.
We will show that the Dirac equation is trivial when the fixed point
equations and the ADHM equations are satisfied.
%
For finite $N$ case, it is known that solutions satisfying the fixed point equations and the ADHM
equations are isolated and classified by the Young diagrams \cite{Nakajima}.
We will give a new proof of this statement 
by solving the ADHM equations and the fixed point equations
concretely and by giving graphical interpretation of the field components
and these equations.

Here is the organization of this article.
In section 2, we review the ${\cal N}= 2$ supersymmetric gauge
theory on ${\mathbb R}^{4}$ and N.C. ${\mathbb R}^{4}$ with a hypermultiplet.
In section 3, a D-brane interpretation is discussed.
In section4, we deform the BRS transformation by using the global symmetries
of the theory.
In section 5, we solve the Seiberg-Witten monopole equations reduced to
$0$ dimension and the fixed point equations, and 
show our main claims.
In section 6, we briefly comment on the localization theorem.
Section 7 is summary of this article.

%
%
\section{${\cal N}= 2$ Supersymmetric $U(1)$ Theory on N.C. ${\mathbb R}^{4}$} \label{n=2theory}
In this section we review ${\cal N}= 2$ supersymmetric 
theory and its topological twist 
on ${\mathbb R}^{4}$ and N.C. ${\mathbb R}^{4}$.
We consider the case with hypermultiplet 
\cite{Hyun-Park-Park1}-\cite{Labastida3}.
For conventions in this article, see appendix \ref{conv}.

At first, we set up the model of the ${\cal N}= 2$ supersymmetric 
theory on ${\mathbb R}^{4}$.
$SO(4)$ spacetime rotation of
$4$ dimensional Euclidean space is locally equivalent to $SU(2)_L \otimes SU(2)_R$.
${\cal N}= 2$ supersymmetric theories have
$SU(2)_I$ R-Symmetry.
The supersymmetry generators $Q_{\alpha i}$, $\bar{Q}_{\dot{\alpha} i}$ 
have indices $i=1,2$ for the R-symmetry.
${\cal N}= 2$ supersymmetric
theories on ${\mathbb R}^{4}$ have following symmetry;
\begin{eqnarray}
H = SU(2)_L \otimes SU(2)_R \otimes SU(2)_I \ .
\end{eqnarray}
The supersymmetric gauge multiplet is given by
\begin{eqnarray}
\begin{array}{ccc}
\quad         &A_{\mu} \quad &{}\\
 \psi^1 \quad       &\phantom{\psi^q}\quad &\psi^2 \ \ .\\
 \quad         &\phi \quad &{}
\end{array}
\end{eqnarray}
Here 
$\psi^1$,$\psi^2$ and $\bar{\psi}^1$,$\bar{\psi}^2$
are Weyl spinors and their CPT conjugate. $\phi$ and
$\bar \phi$ are scalar fields.
Their quantum number of $H$ are assigned as
\begin{eqnarray}
\begin{array}{ccc}
\psi^1 = (1/2,0,1/2),&\psi^2 = (1/2,0,1/2),&\phi = (0,0,0),\\
\bar{\psi}^1=(0,1/2,1/2),&\bar{\psi}^2 = (0,1/2,1/2),&\bar \phi =(0,0,0).
\end{array}
\end{eqnarray}
The action functional is given by
\begin{eqnarray}
L=& -\frac{1}{4}F_{\mu \nu}^{a}F^{\mu \nu}_{a}
-i\bar{\psi}_{\dot{\alpha} i}^{a}
\bar{\sigma}^{\mu \dot{\alpha}\alpha}D_{\mu}\psi_{\alpha a}{}^{i}
-D_{\mu}\bar{\phi}^{a}
D^{\mu}\phi_{a}
\\
& -\frac{i}{\sqrt{2}}\psi^{\alpha ia}[\bar{\phi},\psi_{\alpha i}]_{a}
- \frac{i}{\sqrt{2}}\bar{\psi}_{\dot{\alpha}}{}^{ia}
[\phi,\bar{\psi}^{\dot{\alpha}}{}_{i}]_{a}-
\frac{1}{2}[\bar{\phi},\phi]^2,\ .
\end{eqnarray}
The supersymmetric transformation with parameter $\xi$ and ${\bar \xi}$
are written as
\begin{eqnarray}
\delta A_{\mu}&=&
i\xi^{\alpha i}\sigma_{\mu \alpha\dot{\alpha}}
\bar{\psi}^{\dot{\alpha}}{}_{i}
-i\psi^{\alpha
i}\sigma_{\mu \alpha\dot{\alpha}}\bar{\xi}^{\dot{\alpha}}{}_{i},
\nonumber \\
\delta\psi_{\alpha}{}^{i}&=
&\sigma^{\mu \nu \, \beta}_{\,\,\alpha}\xi_{\beta}{}^{i}F_{\mu \nu}
    +\sqrt{2}i\sigma^{\mu}_{\alpha\dot{\alpha}}D_{\mu}
    \phi \bar{\xi}^{\dot{\alpha} i}
    +[\phi,\bar{\phi}] \xi_{\alpha}{}^{i},
     \nonumber \\
\delta\bar{\psi}_{\dot{\alpha} i} &=
&-\bar{\xi}_{\dot{\beta} i}
\bar{\sigma}^{\mu \nu \dot{\beta}}_{\quad \,\dot{\alpha}}
F_{\mu \nu}
   +\sqrt{2}i\xi^{\alpha i}\sigma^{\mu}_{\alpha \dot{\alpha}}
   D_{\mu}\bar{\phi}
   -[\phi,\bar{\phi}]\bar{\xi}_{\dot{\alpha} i},
     \nonumber \\
\delta \phi&=
&\sqrt{2}\xi^{\alpha i}\psi_{\alpha i},
      \nonumber \\
\delta\bar{\phi}&=
&\sqrt{2}\bar{\xi}^{\dot{\alpha}}_{i}\bar{\psi}_{\dot{\alpha}}{}^{i}.
\end{eqnarray}

To classify the solutions of BPS equations by equivariant cohomology,
let us introduce topological twist here \cite{Witten01,Witten02}.
We use a diagonal
subgroup $SU(2)_{R'}$ in $SU(2)_R \otimes SU(2)_I$ of
$H$. We redefine the spacetime rotation group by
\begin{eqnarray}
K' := SU(2)_L \otimes SU(2)_{R'},
\end{eqnarray}
Then combinations of spinors whose 
quantum number of $H$ are
$(1/2,0,1/2)\oplus (0,1/2,1/2)$
have quantum number $(1/2,1/2)\oplus (0,1)\oplus (0,0)$ of 
$K'$.
Particularly $(0,0)$ is scalar and 
$Q = \epsilon^{\dot{\alpha}i}\bar{Q}_{\dot{\alpha}i}$ is
a BRS operator.
Fermionic fields are similarly topological twisted as
$\psi^i \ (\frac{1}{2} , 0, \frac{1}{2}) 
\rightarrow  \psi_{\mu} \ (\frac{1}{2} , \frac{1}{2})\ 
$
and $
\bar{\psi}^i \ (0,\frac{1}{2} , \frac{1}{2}) 
\rightarrow \chi_{\mu \nu} \ (0,1) \ \oplus \ \eta \ (0,0)$.
BRS transformations are given as
\begin{eqnarray}
{\hat \delta} A_\mu &= i\psi_\mu, \ \ \ \ \ 
{\hat \delta} \psi_\mu &= - D_\mu \phi,\ \ \ \ \ 
{\hat \delta} \phi = 0,
\nonumber \\
{\hat \delta} \chi_{\mu\nu} &= H_{\mu\nu},\ \ \ \ \ 
{\hat \delta} \bar\phi &= i\eta,
\nonumber \\
{\hat \delta} H_{\mu\nu} &= i[\phi,\chi_{\mu\nu}],\ \ \ \ \ 
{\hat \delta} \eta &= [\phi,\bar\phi] \ .
\end{eqnarray}
Here we introduce auxiliary field $H_{\mu \nu}$.

Next step, let us introduce hypermultiplets.
${\cal N}=2$ hypermultiplet consists from two Weyl fermions 
$\lambda_{\!q}$ and $\lambda_{\!\tilde{q}}^\dagger$ and two
complex scalar boson ;
$q$ and $\tilde q^\dagger$
\begin{eqnarray}
 &\lambda_q & \nonumber \\
 q\quad         & &{\tilde{q}}^{\dagger} \ .\nonumber \\
 \quad         &{\lambda}_{\tilde{q}}^\dagger\quad &{}\nonumber
\end{eqnarray}
The definition of the symbol $\dagger$ is seen in appendix \ref{conv}.
Their supersymmetric transformations are given by
\begin{eqnarray}
\delta q^{i}&=
&-\sqrt{2}\xi^{\alpha i}\lambda_{q\alpha}
+\sqrt{2}\bar{\xi}_{\dot{\alpha}}{}^{i}
\bar{\lambda}_{\tilde{q}}^{\dot{\alpha}},
   \nonumber \\
\delta\lambda_{q\alpha}&=
&-\sqrt{2}i\sigma^{\mu}_{\alpha\dot{\alpha}}
   D_{\mu}q^{i}\bar{\xi}^{\dot{\alpha}}{}_{i}
   -2T_{a}q^{i}\bar{\phi}^{a}\xi_{\alpha i},
  \nonumber \\
\delta\bar{\lambda}_{\tilde{q}}^{\dot{\alpha}}&=
&-\sqrt{2} i\bar{\sigma}^{\mu \dot{\alpha}\alpha}
D_{\mu}q^{i}\xi_{\alpha i}
 +2T_{a}q^{i}\phi^{a}\bar{\xi}^{\dot{\alpha}}{}_{i},
\end{eqnarray}
where $T_a$ is a generator of gauge group.
In the following, we consider the case that
representation of the 
gauge group of the hypermultiplet is fundamental
representation.
After topological twisting, 
BRS transformations are given by
%
\begin{eqnarray}
{\hat \delta} q^{\dot\alpha}&=&
 {\psi}_{\!q}^{\dot{\alpha}},\ \ 
{\hat \delta} q_{\dot\alpha}^{\dagger}=
  {\psi}^{\dagger}_{\!q \dot{\alpha}} ,
\nonumber \\
{\hat \delta} {\psi}_{\!q}^{\dot{\alpha}}&=&
-i \phi^{a}T_a q^{\dot{\alpha}},\ \ 
{\hat \delta} {\psi}^{\dagger}_{\!q\dot{\alpha}}=
i q_{\dot\alpha}^{\dagger}\phi^{a}T_{a}, \nonumber \\
{\hat \delta} \chi_{q \alpha} &=& H_{q \alpha},\ \ {\hat \delta}
 \chi^{\dag \alpha}_{q} = H^{\dag \alpha}_{q} \nonumber \\
{\hat \delta} H_{q \alpha} &=& -i \phi^a T_a \chi_{q \alpha},\ \ {\hat
 \delta} H^{\dag \alpha}_{q} = i \chi^{\dag \alpha}_{q} \phi^a T_a, 
\end{eqnarray}
where fields are rescaled 
\footnote{
\begin{eqnarray}
\phi \rightarrow \frac{i}{2\sqrt{2}}\phi, \
\sqrt{2}\bar{\lambda}_{\!q}^{\dot{\alpha}}
\rightarrow \bar{\lambda}_{\!q}^{\dot{\alpha}},\ 
\sqrt{2}\bar{\lambda}_{\!q\dot{\alpha}}
\rightarrow \bar{\lambda}_{\!q\dot{\alpha}},\nonumber
\end{eqnarray}
} and also auxiliary field $H_{q \alpha}$ is
introduced. After topological twisting, we
rename the fermions as $\lambda_q \rightarrow \chi_q $ and 
$\bar{\lambda}_q \rightarrow \psi_q$.\\

Using these field contents, let us construct the action of 
Seiberg-Witten theory.
The action with fundamental hypermultiplet terms are defined by
\begin{eqnarray}
S = k - \hat{\delta} \Psi
\end{eqnarray}
where $k$ is instanton number 
\begin{eqnarray}
k =  \frac{1}{8\pi^2}\int \tr(F_A\wedge F_A),
\end{eqnarray}
and $\Psi$ is a gauge fermion;
\begin{eqnarray}
\Psi 
&=&
-\chi^{\mu\nu a}_+\{
H^{a}_{+\mu\nu }-
s_{+\mu\nu}^a
\}
-\chi^{\dagger\alpha}_q\{
H_{q\alpha}-
s_\alpha
\}
-\{
H^{\dagger\alpha}_q - 
s^{\dagger\alpha}
\}\chi_{q\alpha}
\nonumber
\\
& &
+ i[\phi,{\bar{\phi}}]^a\eta^a
+D_\mu{\bar{\phi}}^a\psi^{\mu a}
-(
-iq^\dagger_{\dot{\alpha}}{\bar{\phi}}
)\psi^{\dot{\alpha}}_q
-\psi^\dagger_{q\dot{\alpha}}(
i{\bar{\phi}}q^{\dot{\alpha}} 
) \ ,
\end{eqnarray}
 Here
 \begin{eqnarray}
s^{\mu \nu}(A,q,q^\dagger) &=& F^{+\mu \nu }_a +q^\dagger\bar\sigma^{\mu \nu }T_a q.
\nonumber \\
s^{\alpha}(A,q) &=& \sigma^\mu  D_\mu  q =\Slash{D}q \ .
\label{mono}
\end{eqnarray}
After integration of the auxiliary fields $H_{+ \mu \nu} $
and $H_q$, the bosonic action are given as
\begin{eqnarray}
S_B
&=& \int\!d^4\!x\sqrt{g}
\biggl[ \frac{1}{4}|s^{\mu \nu}|^2 +\frac{1}{2}|s^{\alpha}|^2\biggl] +
\cdots \ .
\label{sboson}
\end{eqnarray}
Notice that when the gauge group is $U(1)$ and the theory is defined on simple type 
commutative manifolds
we get the Seiberg-Witten invariants as the partition function of this
model \cite{Witten1,Witten2,Hyun-Park-Park1,Hyun-Park-Park2}.
{}From (\ref{sboson}) we get the BPS equations, 
\begin{eqnarray}
s^{\mu \nu}(A,q,q^\dagger) =0 \ , \  s^{\alpha}(A,q)=0  \ , \label{mono_eq}
\end{eqnarray}
which is known as the non-Abelian Seiberg-Witten monopole equations.

In the following, we investigate some 
properties of ${\cal N}= 2$ supersymmetric gauge
theory on N.C. ${\mathbb R}^{4}$
whose noncommutativity is defined as
\begin{eqnarray}
[x^{\mu} , x^{\nu} ] = i \theta^{\mu \nu} \ ,
\end{eqnarray}
where the $\theta^{\mu \nu}$ is an element of an antisymmetric matrix
and called N.C. parameter.
For simplicity, we take 
\begin{eqnarray}
(\theta^{\mu \nu} )=
\left(
\begin{array}{cc|cc}
0& \theta^1 & 0 & 0 \\
-\theta^1 & 0 & 0& 0 \\
\hline
0 & 0& 0& \theta^2 \\
0& 0& -\theta^2 & 0
\end{array}
\right) \ \ . \label{theta1}
\end{eqnarray}

In the following, we only use operator formalisms to describe the
N.C. field theory, therefore
the fields are operators acting on the Hilbert space ${\cal H}$.
Then differential operators $\partial_{\mu}$ are expressed by using
commutation brackets
$ -i \theta^{ -1}_{\mu \nu}[ x^{\nu} , \ *] \equiv [ \hat{\partial}_{\mu} , 
* ]$
and $\int d^{2D} x $ is replaced with $ det (\theta )^{1/2}Tr_{\cal H} $.
\\

When we consider only the case of N.C. ${\mathbb R}^{4}$,
field theories are expressed by the Fock space formalism.
(See appendix in \cite{sako-suzuki}.)
In the Fock space representation,
fields are expressed as $A_{\mu}= \sum {A_{\mu}}^{n_1 n_2}_{m_1 m_2}
|n_1 ,n_2 \rangle \langle m_1 , m_2|$ ,
$\psi_{\mu}= \sum {\psi_{\mu}}^{n_1 n_2}_{m_1 m_2}
|n_1 ,n_2 \rangle \langle m_1 , m_2|$ , etc.
Therefore, the above BRS transformations are expressed as
\begin{eqnarray}
\hat{\delta} {A_{\mu}}^{n_1 n_2}_{m_1 m_2} = {\psi_{\mu}}^{n_1 n_2}_{m_1 
m_2}
\ \ , \ \ \hat{\delta} {\psi_{\mu}}^{n_1 n_2}_{m_1 m_2} = (D_{\mu} \phi )^{n_1 n_2}_{m_1 m_2}\ \ , \ \  \cdots \ \ .
\end{eqnarray}
where the covariant derivative is defined by
$D_{\mu}\  * := [\hat{\partial}_{\mu} + i A_{\mu}\ ,\  * \ ]$
with $\hat{\partial}_{\mu}:= -i \theta^{ -1}_{\mu \nu} x^{\nu} $.
The action functional is given by
\begin{eqnarray}
S&=& Tr_{\cal H} \
L ( A_{\mu}, \dots ;
\hat{\partial}_{z_i}, \hat{\partial}_{\bar{z}_i} ) \nonumber \\
&=& Tr_{\cal H}  tr
 \hat{\delta}\Psi  \ . \label{ctreinta} 
\end{eqnarray}
Let us change the dynamical variables as
\begin{eqnarray}
&&A_{\mu} \rightarrow \frac{1}{\sqrt{\theta}} \tilde{A}_{\mu} , \ \
\psi_{\mu} \rightarrow \frac{1}{\sqrt{\theta}} \tilde{\psi}_{\mu}  , \ \
\bar{\phi} \rightarrow \frac{1}{\theta} \tilde{\bar{\phi}} , \ \
\eta \rightarrow \frac{1}{\theta} \tilde{\eta} , \ \ 
q \rightarrow \frac{1}{\sqrt{\theta}} \tilde{q} , \ \
q^{\dagger} \rightarrow \frac{1}{\sqrt{\theta}} \tilde{q}^{\dagger}
\nonumber \\
&&
\chi_{\mu \nu}^+ \rightarrow \frac{1}{\theta} \tilde{\chi}_{\mu \nu}^+  ,
\ \
H_{\mu \nu}^+ \rightarrow \frac{1}{\theta} \tilde{H}_{\mu \nu}^+ , \ \
\phi \rightarrow \tilde{\phi} \ \  
\psi_q \rightarrow \frac{1}{\sqrt{\theta}} \tilde{\psi_q} , \ \
\psi^{\dagger}_q \rightarrow \frac{1}{\sqrt{\theta}} 
\tilde{\psi}^{\dagger}_q , \nonumber \\
&& 
\chi_q \rightarrow \frac{1}{{\theta}} \tilde{\chi_q} , \ \ 
\chi_q^{\dagger} 
 \rightarrow \frac{1}{{\theta}} \tilde{\chi_q}^{\dagger}
, \ \ 
H_{q} \rightarrow \frac{1}{\theta} \tilde{H}_{q} , \ \
H_{q}^{\dagger} \rightarrow \frac{1}{\theta} \tilde{H}_{q}^{\dagger} 
. \label{weight1} 
\end{eqnarray}
Note that this changing does not cause nontrivial Jacobian
from the path integral measure because of the BRS symmetry.
Then, the action is rewritten as
\begin{eqnarray}
S \rightarrow \frac{1}{\theta^2} \tilde{S} \ \ \ , \ \ \
L( A_{\mu}, \dots ;
\hat{\partial}_{z_i}, \hat{\partial}_{\bar{z}_i} ) \rightarrow 
\frac{1}{\theta^2} L(\tilde{A}_{\mu} , \dots ; -a_i^{\dagger}
, a_i )\ \ .
\end{eqnarray}
Here the action in LHS
depends on $\theta$ because the derivative is given
by $\partial_{z_i}= - \sqrt{\theta^{ -1}} [ a_i^{\dagger} , \ ] $ and so on.
In contrast, the action $\tilde{S}$ in RHS
does not depend on $\theta$ because all $\theta$ parameters are
factorized out.
Using the BRS symmetry, it is proved that
the partition function is invariant under the
deformation of $\theta$, because
 $ \delta_{\theta} Z = -2 (\delta \theta ) \theta^{-3}\langle
\tilde{S} \rangle =0$.
As discussed in \cite{sako-suzuki}, the partition function
of this theory is possible to be determined by using a lower dimension
theory that is given by dimensional reduction.
Therefore, the investigation of the dimensional reduction of the theories is important.

The dimensional reduction of Seiberg-Witten monopole equations (\ref{mono})
are expressed as
\begin{eqnarray}
P^{\mu \nu \rho \tau}_+
[A_{\rho} , A_{\tau} ] + q \bar\sigma^{\mu \nu } q^\dagger =0 \ ,
\label{pre_ADHM} \\
\sigma^\mu  A_\mu  q = 0\ ,
\label{mono_reduction}
\end{eqnarray}
where $P^{\mu \nu \rho \tau}_+$ is a selfdual projection operator.
These expressions are valid for the dimensional reduction of the non-Abelian
theory on commutative ${\mathbb R}^4$.
Using $q_+ := (q_{\dot 1} + q_{\dot 2})/{\sqrt 2}$ and 
$q_- := (q_{\dot 1} - q_{\dot 2})/{\sqrt 2}$, if we start from the $U(1)$ theory
on N.C.${\mathbb R}^4$, the equation (\ref{pre_ADHM}) is 
rewritten as ADHM equations :
\begin{eqnarray}
[ A_{z_1} , A_{z_1}^{\dagger} ] + [ A_{z_2} , A_{z_2}^{\dagger} ]
+ q_- q_-^{*T} -q_+ q_+^{*T} &=& 0 \ , \nonumber \\
{}[A_{z_1} , A_{z_2}] + q_- q_+^{*T} &=& 0 \ . \label{ADHM}
\end{eqnarray}

Note that these operators in (\ref{ADHM}) are expressed by infinite dimensional matrices and the ADHM equations correspond to the instanton of $U(N)$ gauge group with instanton number $N$ at the large $N$ limit.
We consider the finite $N$ situation in the next section.

%
%
\section{D-brane Interpretation} \label{d-brane}
In this article, we study detail of the solution of (\ref{pre_ADHM}) and (\ref{mono_reduction}).
On the N.C. ${\mathbb R}^4$ the fields appearing in (\ref{pre_ADHM}) and (\ref{mono_reduction})
is infinite dimensional matrix acting on Hilbert space.
But the equations are important even if the dimension of the matrix is finite,
because there is a corresponding physical model.
In this section, we consider the correspondence between Seiberg-Witten
monopole equations, D-brane picture and (\ref{pre_ADHM}) (\ref{mono_reduction}) \cite{Popov}.

At first, we construct the physical model
by using the similar manner of the article \cite{Popov}. 
(See also \cite{Popov1}-\cite{Szabo}.)

The generalized second order effective action of $N$ $D3$-brane $N$
$\bar{D}3$-brane
system without topological terms are given by
\begin{eqnarray}
\int tr \left\{ \frac{1}{4}F_{\mu \nu}^{(N)}  F^{(N){\mu \nu}}
+ \frac{1}{4}F_{\mu \nu}^{(\bar{N})}  F^{(\bar{N}){\mu \nu}}
+ | D^{\mu} \phi |^2 + \frac{1}{2}(\tau^2- \phi \bar{\phi})^2
\right\} \ .
\end{eqnarray}
Here the $F_{\mu \nu}^{(N)}$ and $F_{\mu \nu}^{(\bar{N})}$ are
the curvature of the $A^{(N)}$ 
and $A^{(\bar{N})}$ , respectively, where
$A^{(N)}$ 
and $A^{(\bar{N})}$ correspond to open strings attached on $D3$-brane
and $\bar{D}3$-brane.
Up to topological terms, we can rewrite this action as
\begin{eqnarray}
\int tr \left\{ \frac{1}{4}F_{\mu \nu}^{(\bar{N})}  F^{(\bar{N}){\mu \nu}}
+\frac{1}{2}|F_{{z}^1 \bar{z}^1}^{(N)}+F_{{z}^2 \bar{z}^2}^{(N)} 
+(\phi \bar{\phi} - \tau^2)|^2
+8| F_{{z}^1 {z}^2} |^2
+2|D_{\bar{z}_{\bar{1}}}\phi|^2 +2|D_{\bar{z}_{\bar{2}}}\phi|^2
\right\} \ .
\end{eqnarray}

{}From this action, considering the case of $A_{\mu}^{(\bar{N})}=0$,
stationary points are given by
\begin{eqnarray}
F_{{z}^1 \bar{z}^1}^{(N)}+F_{{z}^2 \bar{z}^2}^{(N)} 
+ q_- q_-^{*T} 
&=& \zeta \ \ , \\  
F_{{z}^1 {z}^2}^{(N)} &=& 0 \ \ , \\
D_{\bar{z}^1} q_-  &=& 0 \ \ , \\
D_{\bar{z}^2} q_- &=& 0 \ ,
\end{eqnarray}
where we replace $\phi$ by $q_-$ and $\tau^2$ by $\zeta$. Then, this is
the Seiberg-Witten monopole 
equations with $q_+=0$ condition and back ground constant field $\zeta$.
(See also the next section.) 
This case corresponds to the $\zeta >0$ as we will see in 
section \ref{solution}.
Note that $q_-$ can be regarded as a complex scalar
field when we  consider ${\mathbb R}^4$ case.

The solution of (\ref{pre_ADHM}),(\ref{mono_reduction}) of finite matrix model is 
realized as some $D3$-$\bar{D}3$ configuration. 

%
%
\section{Deformed BRS Transformation} \label{equation}
In this section, we will investigate the symmetry of the dimension reduction of
(\ref{ctreinta}) to $0$ dimension,
and deform the BRS symmetry
as ${\cal G} \otimes T^{N+2}$ equivariant derivative,
where ${\cal G}$ is the gauge transformation group of $U(N)$ and $T^{N+2}$
is the torus action,
in order to derive the fixed point equations.
Note that the $U(N)$ symmetry is caused from the $U(1)$ symmetry if we consider the N.C. theory.
As explained in section \ref{n=2theory}, the action functional is defined by
infinite dimensional matrices when we start from N.C. theories,
then N.C.$U(1)$ gauge symmetry is expressed by $U(\infty)$ symmetry.
%
For simplicity, in some discussions of this paper, 
we restrict our
analysis to the finite dimensional, $N \times N$, matrix case.
( Only proof of the theorem \ref{prop5} in section \ref{solution} and 
the calculations of the 
partition function of a toy model in section \ref{localization}
are based on discussions of finite $N$.) 
All of the fields contents , $A_{\mu} , q $, etc, are given by $N \times N$ matrices.
Then the $U(\infty)$ symmetry is also truncated to $U(N)$.
{}From the viewpoint of N.C.field theory, there might be another type of solutions which is not studied in this article, and the following analysis 
might not be completed.
On the other hand, as discussed in the previous section,
the finite $N \times N$ theory has a $D3$-${\bar D}3$ brane
interpretation, then it has physical applications.

The path integral for cohomological field theories
reduced to the integral over the moduli space of vacuum.
In our case, the moduli space is defined by solutions of
(\ref{pre_ADHM}),(\ref{mono_reduction}).
As demonstrated in \cite{Nekrasov}, 
the localization theorem is a powerful tool for path integrals of
cohomological field theories.
The localization theorem is valid when a theory under consideration has
symmetries under some group actions, and the group actions have isolated
fixed points. 
(For the localization theorem, see also section \ref{localization}.)
Therefore, to investigate solutions of the fixed point equation is
important. This is the main subject of this paper.

Adding to the $U(N)$ gauge symmetry and the Lorentz symmetry
$SO(4)=SU(2)_L \otimes SU(2)_R$,
the action reduced to $0$ dimension has the next extra unitary symmetry, denoted
by ${\tilde U}(N)$,
\begin{equation}
\delta^{{\tilde U}(N)} q_{\dot \alpha} = i q_{\dot \alpha} b,
\label{extra}
\end{equation}
where $b$ is a generator of ${\tilde U}(N)$.\footnote{
When we consider the case that $q_{\dot \alpha}$ is a $N \times k$
matrix in the next section,
then the symmetry becomes ${\tilde U}(k)$;
\begin{equation}
\delta^{{\tilde U}(k)} q_{\dot \alpha} = i q_{\dot \alpha} b \ , \ b \in
 {\tilde u}(k).
\end{equation}
}
Recall that $q$ and $q^{\dagger}$ are fundamental representation of the gauge group.
The gauge transformation of $q$ is defined by left action of the $U(N)$.
Notice that if we define the gauge transformation by using right action,
we can define another gauge symmetry with the corresponding gauge field.
We do not introduce this gauge field, then the symmetry appears only
after the dimensional reduction.
This is the origin of ${\tilde U}(N)$.

Now we use the Abelian subgroup $U(1)^2 \otimes
U(1)^N$ of $SO(4) \otimes {\tilde
U}(N)$. That is, we consider the following symmetry of the action.
\begin{eqnarray}
\delta^{U(1)^2 \otimes U(1)^N} A_{z_i} &=& - i \epsilon_i A_{z_i}, \\
\delta^{U(1)^2 \otimes U(1)^N} q_{\dot \alpha} &=& + i {{M_{R}}_{\dot
 \alpha}}^{\dot \beta} q_{\dot \beta} + i q_{\dot \alpha} b, 
\end{eqnarray}
where
$b = \mbox{diag.} (b_1,\cdots,b_N)$ is a generator of an Abelian
subgroup $U(1)^N$ of ${\tilde U}(N)$, and $\epsilon_i \ (i=1,2)$ is a
generator of an Abelian subgroup $U(1)^2$ of $SO(4)$, defined by
\begin{equation}
\delta A_{\mu} = {M_\mu}^\nu A_\nu \ \ \ , \ \ \ {M_\mu}^\nu = \left(
\begin{array}{cccc}
0 & - \epsilon_1 & & \\
+ \epsilon_1 & 0 & & \\
 & & 0 & - \epsilon_2 \\
 & & - \epsilon_2 & 0
\end{array}
\right).
\end{equation}
Also ${{M_{R}}_{\dot \alpha}}^{\dot \beta}$ is the generator of 
$U(1) \subset SU(2)_R$,
\begin{equation}
{{M_{R}}_{\dot
 \alpha}}^{\dot \beta} = \left(
\begin{array}{cc}
0 & \epsilon_+ \\
\epsilon_+ & 0
\end{array}
\right) \ \ \ , \ \ \ \epsilon_+ = \frac{\epsilon_1 + \epsilon_2}{2}.
\end{equation}

By using above $U(1)^2 \otimes U(1)^N$, let us deform
the BRS symmetry from $\hat{\delta}$ to $\tilde{\delta}$.
We define the deformation by replacing $\hat{\delta}^2=\delta^{U(N) gauge}_{(-\phi)}$ to
\begin{equation}
{\tilde \delta}^{2} = \delta^{U(N) gauge}_{(-\phi)} + \delta^{U(1)^N}_{(b)}+
 \delta^{U(1)^2}_{(\epsilon_{1} , \epsilon_{2})}.
\end{equation}
%
Here $\delta^G_{(\Delta)}$ is a gauge transformation operator with the
group $G$ and the transformation parameter $\Delta$.
Then, for $\psi_{z_{i}}$ and $\psi_{q{\dot \alpha}}$, the BRS transformation rules are
given by,
\begin{eqnarray}
{\tilde \delta}^{2} A_{z_{i}} &=& {\tilde \delta} \psi_{z_{i}} = i [A_{z_{i}} , \phi] - i
 \epsilon_{i} A_{z_{i}}, \label{brsA} \\
{\tilde \delta}^{2} q_{\dot \alpha} &=& {\tilde \delta} \psi_{q {\dot \alpha}} = - i \phi
 q_{\dot \alpha} + {{M_{R}}_{\dot \alpha}}^{\dot \beta} q_{\dot \beta}
 + i q_{\dot \alpha} b, \label{brsq1} \\
{\tilde \delta}^{2} q^{\dag {\dot \alpha}} &=& {\tilde \delta} {\psi_{q}}^{\dag \dot
 \alpha} = q^{\dag {\dot \alpha}} i \phi - {{M_{R}}^{\dot
 \alpha}}_{\dot \beta} q^{\dag {\dot \beta}} - i b q^{\dag {\dot
 \alpha}}. \label{brsq2}
\end{eqnarray}

Now we list the equations, solutions of which we will investigate.
Some of them are the equations of motion,
often called BPS equations.
They are the same as (\ref{pre_ADHM}) or
(\ref{ADHM}),(\ref{mono_reduction}).
However we take some deformation of them, to remove singular solutions.
We introduce a nonzero number $\zeta$, and take
\begin{eqnarray}
& & i ( [A_{z_1} , A_{{\bar z}_1}] + [A_{z_2} , A_{{\bar z}_2}] ) +
q ({\bar \sigma}_{z_1 {\bar z}_1} + {\bar \sigma}_{z_2 {\bar z}_2})
q^{\dag} = i \zeta, \label{eq1} \\
& & i [A_{z_1} , A_{z_2}] + q {\bar \sigma}_{z_1 z_2} q^{\dag} = 0,
 \label{eq2} \\
& & ( A_{z_1} \sigma^{z_1} + A_{{\bar z}_1} \sigma^{{\bar
 z}_1} + A_{z_2} \sigma^{z_2} + A_{{\bar z}_2} \sigma^{{\bar
 z}_2}) q = 0. \label{eq3}
\end{eqnarray}
(\ref{eq1}),(\ref{eq2}) are realized by the redefinition of $s^{\mu
\nu}(A,q,q^\dagger)$
\begin{eqnarray}
 & & s^{\mu \nu}(A,q,q^\dagger) \rightarrow F^{+\mu \nu }
 +q \bar\sigma^{\mu \nu }q^{\dagger} - \zeta^+_{\mu \nu},
\nonumber \\
 & & \zeta_{z_1 {\bar z_1}} + \zeta_{z_2 {\bar z_2}} = i \zeta \ , \ \zeta_{z_1
 z_2} = 0. 
\end{eqnarray}
This constant $\zeta$ is considered as a back ground field and we define its BRS
transformation by $\tilde{\delta} \zeta =0$.
Then, we find that all of the above discussions in previous sections 
are valid although we add this back ground field. 
For later use, we rewrite them into
\begin{eqnarray}
& &  [A_{z_1} , A_{{\bar z}_1}] + [A_{z_2} , A_{{\bar z}_2}] 
-(q_{\dot 2} q_{\dot 1}^{* T} + q_{\dot 1} q_{\dot 2}^{* T}) = \zeta, \label{eq1r} \\
& & [A_{z_1} , A_{z_2}] + \frac{1}{2}(q_{\dot 1} q_{\dot 1}^{* T} -
 q_{\dot 2} q_{\dot 2}^{* T}) + \frac{1}{2}(q_{\dot 1} q_{\dot 2}^{* T} - q_{\dot 2} q_{\dot 1}^{* T}) = 0,
 \label{eq2r} \\
& & (A_{{\bar z}_1} - A_{z_2}) q_{\dot 2}
 - (A_{{\bar z}_1} + A_{z_2}) q_{\dot 1} = 0,
 \label{eq3r1} \\
& & (A_{{\bar z}_2} + A_{z_1}) q_{\dot 2}
 - (A_{{\bar z}_2} - A_{z_1}) q_{\dot 1} = 0.
 \label{eq3r2}
\end{eqnarray}
The rest of the equations to be investigated are
the fixed point equations of the deformed BRS transformation
(\ref{brsA}) - (\ref{brsq2}). They are given
by
\begin{eqnarray}
 & & i [A_{z_{i}} , \phi] - i \epsilon_{i} A_{z_{i}} = 0, \label{eq4} \\
 & & - i \phi q_{\dot \alpha} + {{M_{R}}_{\dot \alpha}}^{\dot \beta}
  q_{\dot \beta} + i q_{\dot \alpha} b = 0. \label{eq5}
\end{eqnarray}

In the next section, we will investigate solutions of
(\ref{eq1}),(\ref{eq2}),(\ref{eq3}),(\ref{eq4}),(\ref{eq5}), and will
show that they have isolated solutions.
This fact guarantees that the localization theorem is valid to our case.

%
%
\section{Solutions of (\ref{eq1}),(\ref{eq2}),(\ref{eq3}),(\ref{eq4}),(\ref{eq5})} \label{solution}
In this section, we solve
(\ref{eq1}),(\ref{eq2})(\ref{eq3}),(\ref{eq4}),(\ref{eq5}), and show
that these equations have only isolated solutions and the solutions are expressed by the Young diagrams.
Notice that our analysis is also valid to a case where $q_{\alpha}$'s
are $N \times k , (k \neq N)$ matrices,
 though we will treat $q_{\alpha}$ as $N \times N$ matrices in this section.
If we take $q_{\dot \alpha}$ to be $N \times k$, $q_{\dot \alpha}^{* T}$
to be $k \times N$ and $b \in u(k)$, our proof in this section includes a new
proof for Prop.5.6. in \cite{Nakajima}.

First of all, we diagonalize $\phi$ by using the $U(N)$ gauge symmetry,
\begin{equation}
 \phi = \mbox{diag.} (\phi_1 , \phi_2 , \cdots , \phi_N).
\end{equation}

Next we tackle (\ref{eq4}) and (\ref{eq5}).
{}From (\ref{eq4}) we see immediately that
if and only if,
\begin{equation}
 \phi_J - \phi_I = \epsilon_i,
\end{equation}
$A_{z_i \ I J}$ could be non-zero,
\begin{equation}
A_{z_i \ I J} \neq 0.
\end{equation}
Also from (\ref{eq5}) we see that
if and only if,
\begin{equation}
 \phi_I = b_J \pm \epsilon_{+},
\end{equation}
$q_{\dot 1 \ I J}$ and $q_{\dot 2 \ I J}$ could be non-zero,
\begin{equation}
 q_{\dot 1 \ I J} = \pm q_{\dot 2 \ I J} \neq 0.
\end{equation}
Notice $q_{\dot 1 \ I J}$ and $q_{\dot 2 \ I J}$ are not independent
from one another.

These observations lead us to the following proposition.
\begin{lemma}
If (\ref{eq1}),(\ref{eq4}),(\ref{eq5}) have a solution,
then
$\phi_{I}$ takes any of $\varphi_{[x_{\hat I}]}^{(n_1,n_2)}$, given by 
\begin{equation}
\varphi_{[x_{\hat I}]}^{(n_1,n_2)} = x_{\hat I} + n_1 \epsilon_1 + n_2
 \epsilon_2 \ \ \ , \ \ \ n_1 , n_2 \in {\mathbb Z} 
\end{equation}
where
\begin{equation}
x_{\hat I} \in \{ b_{I}^{(-)} \in {\mathbb R} , I = 1,\cdots,N |
 b_{I}^{(-)} := b_{I} - \epsilon_+ \},
\end{equation}
or
\begin{equation}
x_{\hat I} \in \{ y_{\bar I} \in {\mathbb R} , {\bar I} = 1,\cdots,{\bar
 N} | \forall I,n_1,n_2, \ y_{\bar I} \neq b_{I}^{(-)} + n_1 \epsilon_1
 + n_2 \epsilon_2 \}.
\end{equation}
\label{prop1}
\end{lemma}
(proof)\\
Suppose that $\phi_I$ does not take any of $\varphi_{[x_{\hat
I}]}^{(n_1,n_2)}$ given above. This implies that 
$ \exists I, \forall J, \ q_{\dot \alpha \ I J} = 0, A_{z_i \ I J} = A_{z_i \ J I} =
0$.
Consider (\ref{eq1}).
It is easy to see that the $(I,I)$ component of
LHS of (\ref{eq1}) is $0$,
whereas the $(I,I)$ component of RHS of (\ref{eq1}) is $ i
\zeta \neq 0$. 
Therefore no solution to (\ref{eq1}),(\ref{eq4}),(\ref{eq5}) is allowed. $\blacksquare$

\begin{figure}
 \centering
 \includegraphics[width=40mm,clip]{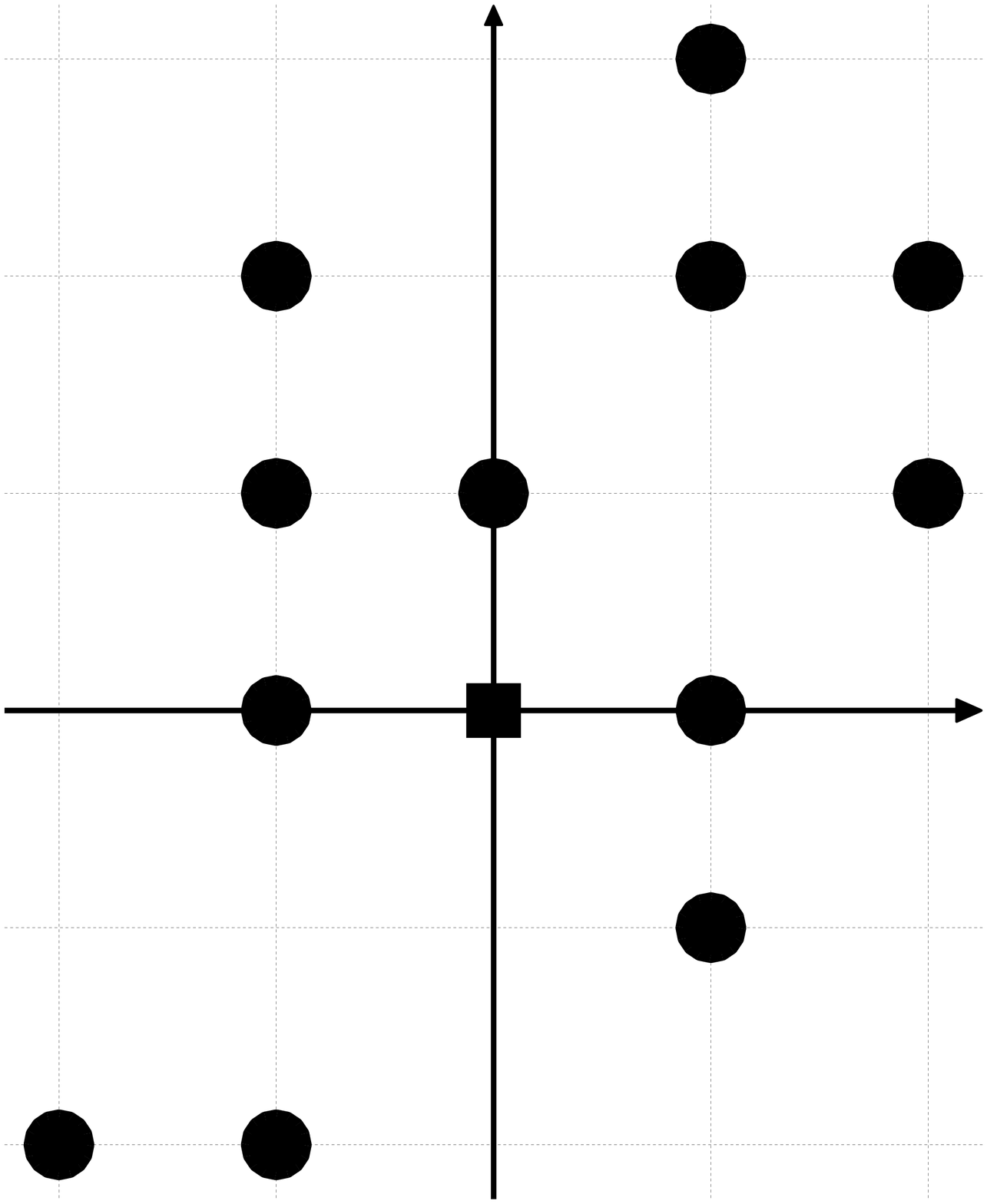}
 \caption{ $P_{[{\hat x}_I]}$ }
 \label{f1}
\end{figure}

For a set of all $\{\varphi_{[x_{\hat I}]}^{(n_1,n_2)} | x_{\hat I} \ \mbox{is given} \}$, assign a graph $P_{[x_{\hat I}]}$. 
See Fig.\ref{f1}. 
In Fig.\ref{f1}, the origin, denoted by the black square, corresponds to the eigenvalue
$\varphi_{[x_{\hat I}]}^{(0,0)} = x_{\hat I}$,
and other lattice points $(n_1,n_2)$, denoted by black dots,
correspond to eigenvalues $\varphi_{[x_{\hat I}]}^{(n_1,n_2)}$. 
For given a set of $P_{[x_{\hat I}]}$,
$\phi$ is written as
\begin{eqnarray}
\phi &=& \bigoplus_{I} \left(
\begin{array}{cccc}
\varphi_{[b_{I}^{(-)}]}^{(n_1,n_2)} {\bf
 1}_{N_{[b_{I}^{(-)}]}^{(n_1,n_2)}} & & & \\
 & \varphi_{[b_{I}^{(-)}]}^{(n'_1,n'_2)} {\bf
 1}_{N_{[b_{I}^{(-)}]}^{(n'_1,n'_2)}}& & \\
 & & \varphi_{[b_{I}^{(-)}]}^{(n''_1,n''_2)} {\bf
 1}_{N_{[b_{I}^{(-)}]}^{(n''_1,n''_2)}}& \\
 & & & \ddots
\end{array}
\right) \\
& & \bigoplus_{\bar I} \left(
\begin{array}{cccc}
\varphi_{[y_{\bar I}]}^{(n_1,n_2)} {\bf
 1}_{N_{[y_{\bar I}]}^{(n_1,n_2)}} & & & \\
 & \varphi_{[y_{\bar I}]}^{(n'_1,n'_2)} {\bf
 1}_{N_{[y_{\bar I}]}^{(n'_1,n'_2)}}& & \\
 & & \varphi_{[y_{\bar I}]}^{(n''_1,n''_2)} {\bf
 1}_{N_{[y_{\bar I}]}^{(n''_1,n''_2)}}& \\
 & & & \ddots 
\end{array}
\right).
\label{phi1}
\end{eqnarray}
In each ${I}$-th or ${\bar I}$-th block, we suppose that eigenvalues
$\varphi_{[b_{I}^{(-)}]}^{(n_1,n_2)}$ or $\varphi_{[y_{\bar I}]}^{(n_1,n_2)}$ are arranged by order,
\begin{eqnarray}
 & & 
\varphi_{[b_{I}^{(-)}]}^{(n_1,n_2)} <
\varphi_{[b_{I}^{(-)}]}^{(n'_1,n'_2)} <
\varphi_{[b_{I}^{(-)}]}^{(n''_1,n''_2)} < \cdots, \nonumber \\
 & & 
\varphi_{[y_{\bar I}]}^{(n_1,n_2)} < \varphi_{[y_{\bar
I}]}^{(n'_1,n'_2)} < \varphi_{[y_{\bar I}]}^{(n''_1,n''_2)} < \cdots.
\label{order}
\end{eqnarray}

The index $I$ is mapped to the triad of indices $( {\hat I},(n_1,n_2) )$,
\begin{equation}
I \mapsto ( {\hat I},(n_1,n_2) ).
\end{equation}
We denote the degeneracy of $\varphi_{[x_{\hat
I}]}^{(n_1,n_2)}$ as $N_{[x_{\hat I}]}^{(n_1,n_2)}$,
\begin{equation}
{}^{\#} \{ \phi_I | \phi_I = \varphi_{[x_{\hat I}]}^{(n_1,n_2)} \} = N_{[x_{\hat
 I}]}^{(n_1,n_2)} \geq 0,
\end{equation} 
\begin{equation}
\sum_{\hat I} \sum_{(n_1,n_2)} N_{[x_{\hat I}]}^{(n_1,n_2)} = N.
\end{equation}

\begin{figure}
 \centering
 \includegraphics[width=40mm,clip]{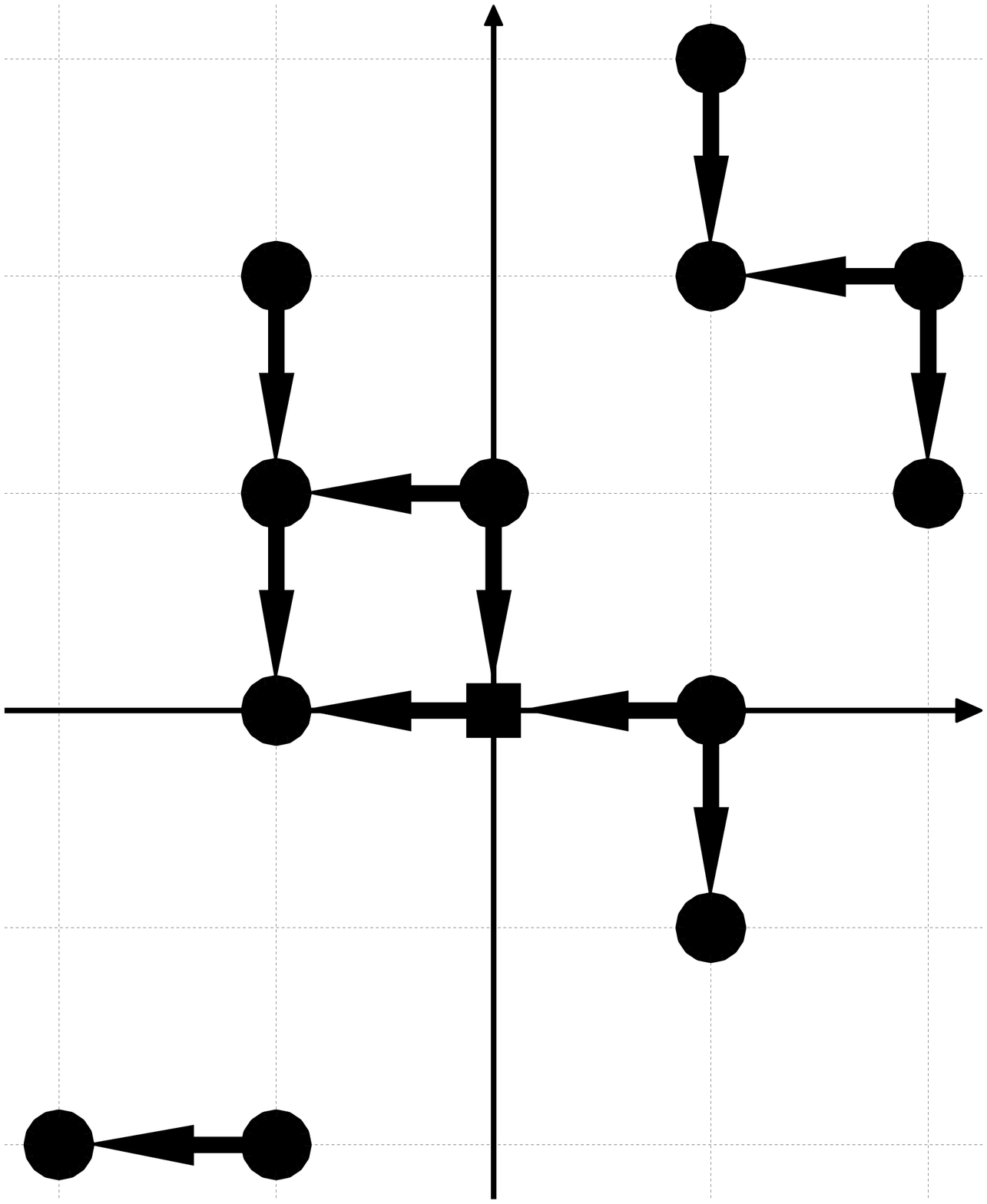}
 \caption{ $G_{[{\hat x}_I]}$ }
 \label{f2}
\end{figure}

$A_{z_i}$ takes a similar block structure,
\begin{eqnarray}
 A_{z_i} &=& \bigoplus_{I} \left(
\begin{array}{ccc}
 & \vdots & \\
\cdots & A_{z_i \ (I,(n_1,n_2)),(I,(m_1,m_2))} & \cdots \\
 & \vdots & 
\end{array}
\right) \nonumber \\
 & & \bigoplus_{\bar I} \left(
\begin{array}{ccc}
 & \vdots & \\
\cdots & E_{z_i \ ({\bar I},(n_1,n_2)),({\bar I},(m_1,m_2))} & \cdots \\
 & \vdots & 
\end{array}
\right),
\label{A1}
\end{eqnarray}
where 
\begin{eqnarray}
 A_{z_i \ (I,(n_1,n_2)),(I,(m_1,m_2))} && \mbox{ is a }
 N_{[b_I^{(-)}]}^{(n_1,n_2)} \times N_{[b_I^{(-)}]}^{(m_1,m_2)} \ \mbox{complex matrix , and }
 \nonumber \\
 E_{z_i \ ({\bar I},(n_1,n_2)),({\bar I},(m_1,m_2))} && \mbox{ is a }
 N_{[y_{\bar I}]}^{(n_1,n_2)} \times N_{[y_{\bar I}]}^{(m_1,m_2)} \ \mbox{complex matrix}.  \nonumber
\end{eqnarray}
A non-trivial component of $A_{z_1}$ appears in 
$\{ ({\hat I},(n_1,n_2)) \ , \ ({\hat I},(n_1 -1,n_2)) \}$-th block 
and, that of $A_{z_2}$ appears in  
$\{ ({\hat I},(n_1,n_2)) \ , \ ({\hat I},(n_1,n_2 -1)) \}$-th block,
\begin{eqnarray}
A_{z_1 \ (I,(n_1,n_2)),(I,(n_1-1,n_2))} \neq 0  &,& 
E_{z_1 \ ({\bar I},(n_1,n_2)),({\bar I},(n_1-1,n_2))} \neq 0 \\
A_{z_2 \ (I,(n_1,n_2)),(I,(n_1,n_2-1))} \neq 0  &,& 
E_{z_2 \ ({\bar I},(n_1,n_2)),({\bar I},(n_1,n_2-1))} \neq 0 \ .
\end{eqnarray}
By adding left-arrows connecting $(n_1,n_2)$ and $(n_1 -1,n_2)$
and down-arrows connecting $(n_1,n_2)$ and $(n_1,n_2 -1)$ to
the graph $P_{[x_{\hat I}]}$,
we obtain a graph $G_{[x_{\hat I}]}$. For example, see
Fig.\ref{f2}. The left-arrow corresponds to $A_{z_1}$'s non-trivial
component, and the down-arrow corresponds to $A_{z_2}$'s non-trivial
component. 
Also the non-trivial components of $q_{\dot \alpha}$ are
\begin{eqnarray}
q_{{\dot 1} \ (I , (0,0)) , J } = - q_{{\dot 2} \ (I ,
  (0,0)) , J } \neq 0 &,& \mbox{for} \ I,J, \ \mbox{s.t.} \ \phi_I = b_J
  + \epsilon_+ ,
\label{q1a} \\
q_{{\dot 1} \ (I , (1,1)) , J } = + q_{{\dot 2} \ (I ,
  (1,1)) , J } \neq 0 &,& \mbox{for} \ I,J, \ \mbox{s.t.} \ \phi_I = b_J - \epsilon_+ .
\label{q1b}
\end{eqnarray}

{}From (\ref{A1}),(\ref{q1a}),(\ref{q1b}), we obtain
the next proposition.
\begin{lemma}
If $\phi_I$ takes any of $\varphi_{[{y_{\bar I}}]}^{(n_1,n_2)} = y_{\bar
 I} + n_1 \epsilon_1 + n_2 \epsilon_2$,
then (\ref{eq1}),(\ref{eq4}),(\ref{eq5}) have no solution.
\label{prop2}
\end{lemma}
(proof) \\
Suppose that some $\phi_I$ are given
by
\begin{equation}
\phi_I = \varphi_{[y_{\bar I}]}^{(n_1,n_2)}.
\end{equation}
Then, LHS of (\ref{eq1r}), equivalent to (\ref{eq1}),
is given by 
\begin{eqnarray}
 & & \mbox{LHS of (\ref{eq1r})} = \sum_{i=1,2} [A_{z_i} , A_{{\bar z}_i}] - (q_{\dot 2} q_{\dot 1}^{*
  T} + q_{\dot 1} q_{\dot 2}^{* T}) \nonumber \\
 &=&  
\left(
\begin{array}{cc}
\bigoplus_{I} \sum_{i=1,2} [A^{I}_{z_i} , A^{I}_{{\bar z}_i}] - (q_{\dot
 2} q_{\dot 1}^{* T} + q_{\dot 1} q_{\dot 2}^{* T}) & 0 \\
0 & \bigoplus_{\bar I} \sum_{i=1,2} [E^{\bar I}_{z_i} , E^{\bar I}_{{\bar z}_i}]
\end{array}
\right),
\label{lhseq1r}
\end{eqnarray}
because the non-trivial components of $q_{\dot \alpha}$ are given by
 (\ref{q1a}),(\ref{q1b}).
On the other hand, RHS of (\ref{eq1r}) is proportional to a unit matrix,
\begin{equation}
\mbox{RHS of (\ref{eq1r})} = \zeta 
\left(
\begin{array}{cc}
\bigoplus_{I} {\bf 1}^{I,I} & 0 \\
0 & \bigoplus_{\bar I} {\bf 1}^{{\bar I},{\bar I}}
\end{array}
\right).
\label{rhseq1r}
\end{equation}
The $({\bar I} \ , \ {\bar I})$ block of (\ref{lhseq1r}) is a traceless matrix,
whereas the $({\bar I} \ , \ {\bar I})$ block of (\ref{rhseq1r}) has a non-zero
trace.
These are mutually exclusive. $\blacksquare$ \\
When we consider the case of $N=\infty$, we can not use the 
nature that the commutator is traceless, then
this proof is not correct. But we can prove this statement
even if $N=\infty$. Because, if $[E^{\bar I}_{z_i} , E^{\bar I}_{{\bar z}_i}]$
is not traceless, we can show that the curvature $F$ does not converge
to zero at infinity. 
This means that if the set of the gauge fields is
$\{A | \lim_{x\rightarrow \infty}|F(x)|=0 \}$,
then this theorem still holds. 
By the same reason, the theorem \ref{prop4} in this section is valid
for $N=\infty$ case.
That is why, all theorems in this section 
without the theorem \ref{prop5}
holds for $N=\infty$ case. 

\begin{corollary}
(\ref{eq1}),(\ref{eq4}),(\ref{eq5}) can have a solution,
if and only if $\phi$ is given by
\begin{equation}
\phi = \bigoplus_{I} \bigoplus_{(n_1,n_2) \in G_{[b^{(-)}_{I}]}} \varphi_{[b_{I}^{(-)}]}^{(n_1,n_2)} {\bf
 1}_{N_{[b^{(-)}_{I}]}^{(n_1,n_2)}},
\label{phi2}
\end{equation}
\begin{equation}
\varphi_{[b_{I}^{(-)}]}^{(n_1,n_2)} = b^{(-)}_{I} + n_1 \epsilon_1 + n_2
 \epsilon_2,
\label{barphibI} 
\end{equation}
and $A_{z_i}$ is given by
\begin{equation}
 A_{z_i} = \bigoplus_{I} A^{I}_{z_i}.
\label{A2}
\end{equation}
\label{prop3}
\end{corollary}

\begin{figure}
 \centering
 \includegraphics[width=40mm,clip]{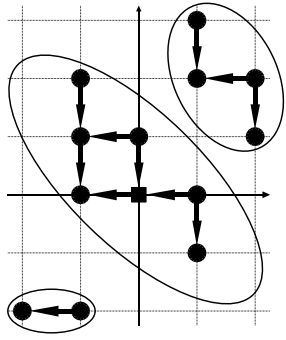}
 \caption{ $G_{[b_I^{(-)}]}$ consists of connected graphs $G_{[b_I^{(-)}]}^{(a)}$ }
 \label{f3}
\end{figure}

{}From now on, we suppose that the parameter $\zeta$ is a positive number,
\begin{equation}
\zeta > 0.
\label{zeta>0}
\end{equation}
(If we assume $\zeta < 0$, we have to change some statements
in the following theorems, but essentially same theorems hold.)
Then we obtain the next theorem.
\begin{theorem}
Let $G_{[b_{I}^{(-)}]}$ be a graph defined from the eigenvalues
 $\varphi_{[b_{I}^{(-)}]}^{(n_1,n_2)}$ given by (\ref{phi2}).
Also let $\zeta$ be positive.
The following three conditions
are necessary
for a solution of (\ref{eq1}),(\ref{eq4}) and (\ref{eq5}) to exist. \\
(1) $G_{[b_{I}^{(-)}]}$ consists of one connected part. \\
(2) $G_{[b_{I}^{(-)}]}$ includes the origin $(0,0)$. \\
(3) All points $(n_1,n_2)$ in $G_{[b_{I}^{(-)}]}$ must be in $n_1 \leq 0 \ , \ n_2 \leq 0$.
\label{prop4}
\end{theorem}
(proof) \\
First of all, notice that $A^I_{z_i}$ is a direct sum of upper triangle (block) matrices and $A^I_{{\bar
z}_i}$ is of lower triangle (block) matrices, (remember (\ref{order}),)
\begin{eqnarray}
 & &  A^I_{z_i} = \bigoplus_{a} A^{I \ (a)}_{z_i} = \bigoplus_{a} \left(
\begin{array}{ccccc}
0 & * & \cdots & * & * \\
0 & 0 & \cdots & * & * \\
\vdots & \vdots & \ddots & \vdots & \vdots \\
0 & 0 & \cdots & 0 & * \\
0 & 0 & \cdots & 0 & 0
\end{array}
\right),
\label{utri} \\
 & & A^I_{{\bar z}_i} = \bigoplus_{a} A^{I \ (a)}_{{\bar z}_i} = \bigoplus_{a} \left(
\begin{array}{ccccc}
0 & 0 & \cdots & 0 & 0 \\
* & 0 & \cdots & 0 & 0 \\
\vdots & \vdots & \ddots & \vdots & \vdots \\
* & * & \cdots & 0 & 0 \\
* & * & \cdots & * & 0
\end{array}
\right),
\label{dtri}
\end{eqnarray}
where the index $a$ labels connected diagrams $G_{[b_I^{(-)}]}^{(a)}$ in $G_{[b_I^{(-)}]}$. See Fig.\ref{f3}.
{}From (\ref{utri}) and (\ref{dtri}), we obtain
\begin{equation}
[A^{I \ (a)}_{z_i} , A^{I \ (a)}_{{\bar z}_i}] =
\left(
\begin{array}{ccc}
M_{min} & * & 0 \\
* & M_{int} & * \\
0 & * & M_{max}
\end{array}
\right), \label{AA2} 
\end{equation}
where
\begin{eqnarray}
 & & M_{min} = + \sum_{(m_1,m_2)} A^{I \ (a)}_{z_i \ (n_1^{min},n_2^{min}) ,
 (m_1,m_2) } A^{I \ (a)}_{{\bar z}_i \ (m_1,m_2)
 , (n_1^{min},n_2^{min})}, \label{AA2min} \\
 & & M_{max} = - \sum_{(m_1,m_2)} A^{I \ (a)}_{{\bar z}_i \
      (n_1^{max},n_2^{max}),(m_1,m_2)} A^{I \ (a)}_{z_i \
      (m_1,m_2),(n_1^{max},n_2^{max})}, \label{AA2max}
\end{eqnarray}
and
\begin{equation}
M_{int} = \left(
\begin{array}{cccc}
M_{int}^{(n_1,n_2)} & * & * & \\
* & M_{int}^{(n'_1,n'_2)} & * & \\
* & * & M_{int}^{(n''_1,n''_2)} & \\
 & & & \ddots 
\end{array}
\right), \label{AA2int}
\end{equation}
\begin{eqnarray}
M_{int}^{(n_1,n_2)} &=&
 + \sum_{(m_1,m_2)} A^{I \ (a)}_{z_i \ (n_1,n_2),(m_1,m_2)} A^{I \ (a)}_{{\bar z}_i \
  (m_1,m_2),(n_1,n_2)} \nonumber \\ 
 & & - \sum_{(m_1,m_2)} A^{I \ (a)}_{{\bar z}_i \
  (n_1,n_2),(m_1,m_2)} A^{I \ (a)}_{z_i \ (m_1,m_2),(n_1,n_2)},
  \nonumber \\
 & &  \ \ \ \cdots \ \ \ .
 \label{AA2intn}
\end{eqnarray}
$(n_1^{min},n_2^{min})$ in (\ref{AA2min}) denotes the point corresponding to the
lowest eigenvalue in $G_{[b_I^{(-)}]}^{(a)}$,
and
$(n_1^{max},n_2^{max})$ in (\ref{AA2max}) denotes the point corresponding to the
highest eigenvalue in $G_{[b_I^{(-)}]}^{(a)}$.
Also
$(n_1,n_2), \cdots $ in (\ref{AA2int}) denote other points corresponding to
intermediate eigenvalues in $G_{[b_I^{(-)}]}^{(a)}$. 
Let us consider a $\{ (I \ (a)) \ , \ (I \ (a)) \}$ block of (\ref{eq1r}),
\begin{equation}
\sum_{i=1,2} [A^{I \ (a)}_{z_i} , A^{I \ (a)}_{{\bar z}_i}] 
-(q_{\dot 2} q_{\dot 1}^{* T} + q_{\dot 1} q_{\dot 2}^{* T})_{ \{ (I \
(a)) \ , \ (I \ (a)) \} } = \zeta \ {\bf 1}_{ \{ (I \ (a)) \ , \ (I \
 (a)) \} }.
\label{eq1rIa}
\end{equation}
If a connected part $G^{(a)}_{[b_I^{(-)}]}$ does not include $(0,0)$ or
$(1,1)$, the second term in LHS of (\ref{eq1rIa}) vanishes, since the
non-trivial components of $q_{\dot \alpha}$ are given by (\ref{q1a}),(\ref{q1b}).
We have supposed $\zeta > 0$, so (\ref{AA2})-(\ref{AA2intn}) tell us that
such $G^{(a)}_{[b_I^{(-)}]}$ does not exist.

Next, consider the $\{(I,(n_1^{max},n_2^{max})) \ , \
(I,(n_1^{max},n_2^{max})) \}$
block of (\ref{eq1r}),
\begin{eqnarray}
 & & - \sum_{(m_1,m_2)} A^I_{{\bar z}_i \
      (n_1^{max},n_2^{max}),(m_1,m_2)} A^I_{z_i \
      (m_1,m_2),(n_1^{max},n_2^{max})} \nonumber \\
 & & -(q_{\dot 2}
 q_{\dot 1}^{* T} + q_{\dot 1} q_{\dot 2}^{* T})_{\{ (I,(n_1^{max},n_2^{max})) \ , \ (I,(n_1^{max},n_2^{max})) \}} \nonumber \\
 &=& \zeta \ {\bf 1}_{N^{(n_1^{max},n_2^{max})}_{[b^{(-)}_I]}}.
\label{eq1rmax}
\end{eqnarray}
If 
\begin{equation}
n_1^{max} > 1 \ \ \ \mbox{or} \ \ \ n_2^{max} > 1,
\end{equation}
the second term in LHS of (\ref{eq1rmax}) vanishes,
since the non-trivial components of $q_{\dot \alpha}$ are given by (\ref{q1a}),(\ref{q1b}),
then  
\begin{equation}
\mbox{LHS of (\ref{eq1rmax})} = - \sum_{(m_1,m_2)} A^I_{{\bar z}_i \
      (n_1^{max},n_2^{max}),(m_1,m_2)} A^I_{z_i \
      (m_1,m_2),(n_1^{max},n_2^{max})} \leq 0.
\end{equation}
On the other hand, 
\begin{equation}
\mbox{RHS of (\ref{eq1rmax})} = \zeta > 0.
\end{equation}
These are inconsistent from each other.
Then, we conclude
\begin{equation}
n_1^{max} \leq 1 \ \ \ \mbox{and} \ \ \ n_2^{max} \leq 1.
\end{equation}

Consider the maximal case, the $\{ (I , (1,1)) \ , \ (I , (1,1)) \}$ component of
(\ref{eq1r}).
The first term in LHS is
\begin{equation}
- \sum_{(m_1,m_2)} A^I_{{\bar z}_i \
      (1,1),(m_1,m_2)} A^I_{z_i \
      (m_1,m_2),(1,1)} \leq 0,
\end{equation}
and the second term is
\begin{equation}
-(q_{\dot 2} q_{\dot 1}^{* T} + q_{\dot 1} q_{\dot 2}^{* T})
 = - 2 q_{\dot 1} q_{\dot 1}^{* T} \leq 0 .
\end{equation}
Again, RHS is $\zeta > 0$. Then we see that the $\{ I (1,1) \}$
component does not exist.
Repeating similar arguments, we conclude that
\begin{equation}
(n_1^{max},n_2^{max}) = (0,0).
\end{equation}
We have finished the proof of Theorem\ref{prop4}. $\blacksquare$

\begin{figure}
 \centering
 \includegraphics[width=30mm,clip]{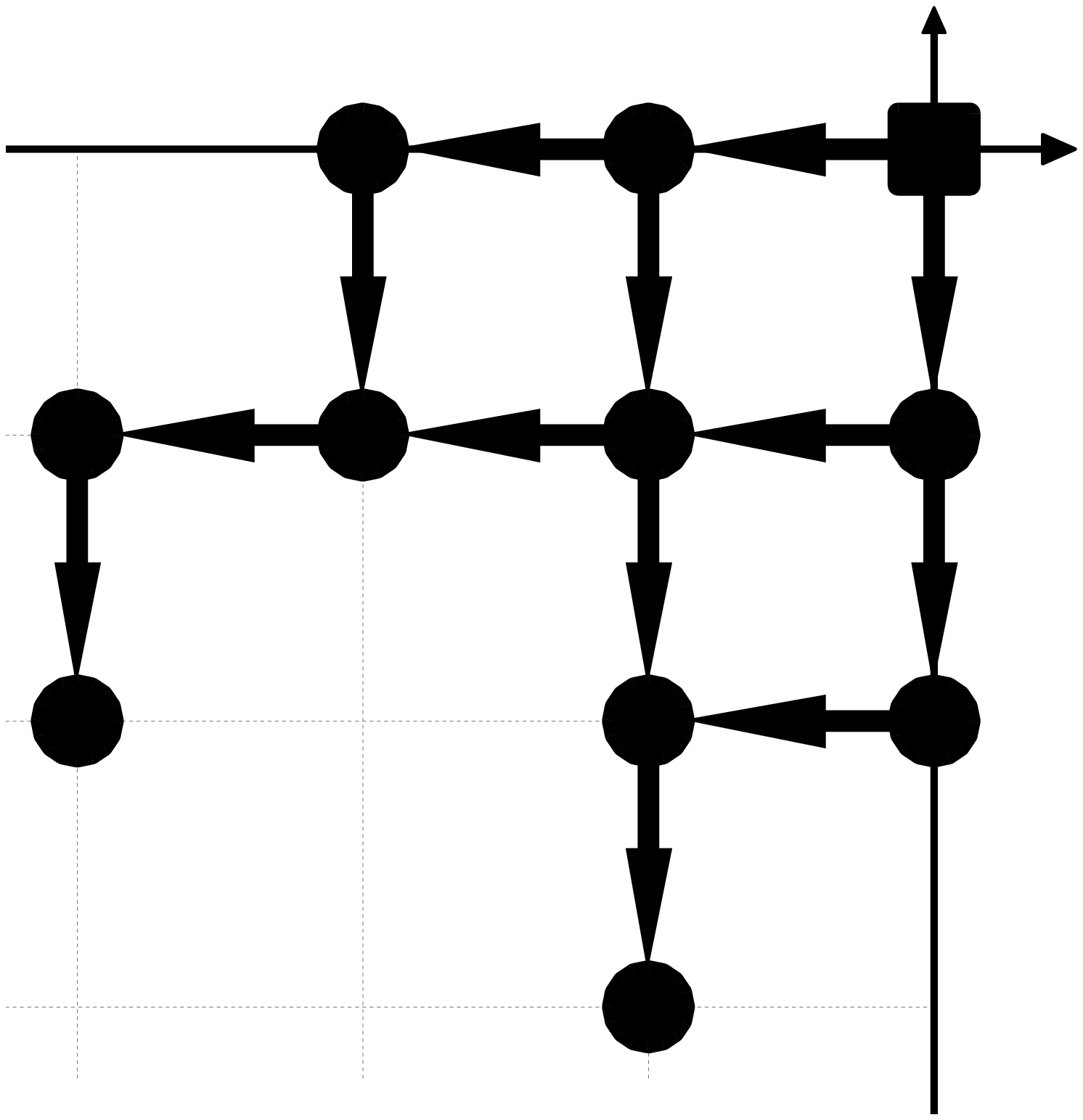}
 \caption{ $C_{{\cal I}(l)}$ }
 \label{f4}
\end{figure}

Let us introduce such a map ${\cal I}$, that
\begin{eqnarray}
& & {\cal I} \ : \ \{ \ l \ | \ l = 1 , \cdots , M \ \} \rightarrow \{
  \ I \ | \ I = 1 , \cdots , N \ \} \ , \ M \leq N, \\
& & N_{[b_{{\cal I}(l)}]}^{(0,0)} \neq 0.
\end{eqnarray}
For each $l$, assign a connected graph $C_{{\cal I}(l)}$.
For example, see Fig.\ref{f4}.
For given $C_{{\cal I}(l)}$, non-trivial components of $A_{z_i}$ are
\begin{equation}
A_{z_1 \ \{ l , (n_1 - 1 , n_2) \} \{ l , (n_1 , n_2) \} } \neq 0 \ \ \
 , \ \ \ (n_1 - 1 , n_2),(n_1 , n_2) \in C_{{\cal I}(l)},
\label{a1n}
\end{equation}
and
\begin{equation}
A_{z_2 \ \{ l , (n_1 , n_2 - 1) \} \{ l , (n_1 , n_2) \} } \neq 0 \ \ \
 , \ \ \ (n_1 , n_2 - 1),(n_1 , n_2) \in C_{{\cal I}(l)}.
\label{a2n}
\end{equation}
Also non-trivial components of $q_{\dot \alpha}$ are
\begin{equation}
 q_{\dot 1 \ I = \{ l , (0,0) \},J = {\cal I}(l)} = - q_{\dot 2 \ I = \{
  l , (0,0) \},J = {\cal I}(l)} \neq 0.
\label{qn}
\end{equation}

For the non-trivial components (\ref{a1n}) - (\ref{qn}),
(\ref{eq1}) and (\ref{eq2}) are reduced to
\begin{eqnarray}
 & & { \ \ } A_{z_1} {}_{ \{ l , (n_1,n_2) \} , \{ l , (n_1+1,n_2)\} }
  A_{{\bar z}_1} {}_{ \{ l , (n_1+1,n_2) \} , \{ l , (n_1,n_2)\} }
  \nonumber \\
 & & - A_{{\bar z}_1} {}_{ \{ l , (n_1,n_2) \} , \{ l , (n_1-1,n_2)\} }
  A_{z_1} {}_{ \{ l , (n_1-1,n_2) \} , \{ l , (n_1,n_2)\} } \nonumber \\
 & & + A_{z_2} {}_{ \{ l , (n_1,n_2) \} , \{ l , (n_1,n_2+1)\} }
  A_{{\bar z}_2} {}_{ \{ l , (n_1,n_2+1) \} , \{ l , (n_1,n_2)\} }
  \nonumber \\
 & & - A_{{\bar z}_2} {}_{ \{ l , (n_1,n_2) \} , \{ l , (n_1,n_2-1)\} }
  A_{z_2} {}_{ \{ l , (n_1,n_2-1) \} , \{ l , (n_1,n_2)\} } \nonumber \\
 & & + 2 q_{\dot 1} {}_{\{l,(n_1,n_2)\},J} \ q_{\dot 1}^{* T}
  {}_{J,\{l,(n_1,n_2)\}} \nonumber \\
 & & = \zeta,
\label{aarn}
\end{eqnarray}
and
\begin{eqnarray}
 & & { \ \ } A_{z_1} {}_{ \{ l , (n_1,n_2) \} , \{ l , (n_1+1,n_2)\} }
  A_{z_2} {}_{ \{ l , (n_1+1,n_2) \} , \{ l , (n_1+1,n_2+1)\} }
  \nonumber \\
 & & - A_{z_2} {}_{ \{ l , (n_1,n_2) \} , \{ l , (n_1,n_2+1)\} } A_{z_1}
  {}_{ \{ l , (n_1,n_2+1) \} , \{ l , (n_1+1,n_2+1)\} } \nonumber \\
 & & = 0.
\label{aacn}
\end{eqnarray}

On the other hand, the Dirac equation reduced to $0$ dimension
(\ref{eq3}) gives no constraint,
which follows from the next theorem.
\begin{theorem}
If $A_{z_i}$ and $q_{\dot \alpha}$ satisfy eqs.(\ref{eq1}),(\ref{eq2})
 and eqs.(\ref{eq4}),(\ref{eq5}), they satisfy 
 the Dirac equation reduced to $0$
 dimension (\ref{eq3}) automatically. 
\label{propD}
\end{theorem}
(proof) \\
{}From (\ref{qn}), (\ref{eq3}) is reduced to
\begin{equation}
A_{{\bar z}_1} q_{\dot 1} = 0 \ , \ A_{{\bar z}_2} q_{\dot 1} = 0.
\label{eqq1n}
\end{equation}
Since we have taken the ordering (\ref{order}),
$A_{{\bar z}_i \ (l,(n_1,n_2)),(l,(m_1,m_2))}$ and $q_{{\dot 1} \
(l,(n_1,n_2)),J={\cal I}(l)}$ have the next structures,
\begin{equation}
A_{{\bar z}_i \ (l,(n_1,n_2)),(l,(m_1,m_2))} = 
\left(
\begin{array}{ccccc}
0 & 0 & \cdots & 0 & 0 \\
* & 0 & \cdots & 0 & 0 \\
\vdots & \vdots & \ddots & \vdots & \vdots \\
* & * & \cdots & 0 & 0 \\
* & * & \cdots & * & 0
\end{array}
\right) \ , \ q_{{\dot 1} \ (l,(n_1,n_2)),J={\cal I}(l)} = 
\left(
\begin{array}{c}
0 \\
\vdots \\
0 \\
*
\end{array}
\right).
\end{equation}
So, (\ref{eqq1n}) always holds. $\blacksquare$

The above theorem means that the solutions of 
the dimensional reduction of the Seiberg-Witten monopole equations with the constant back ground under the fixed point conditions of the torus actions are equivalent to the solutions of the N.C.ADHM equations with the same fixed point conditions.

The above discussions and theorems are valid for 
infinite $N$ as well as finite $N$. 
In the following, we consider only a finite $N$ case
to study more details.
As we saw in section \ref{d-brane}, the finite $N$ case
itself has a physical picture.
Furthermore, solutions and their
natures of finite $N$ models are important even if we consider
the N.C. field theory, because such solutions are possible
to be embedded in infinite $N$ solutions.

%
{}From now on, we suppose that $\phi_I$ does not degenerate,
\begin{equation}
N_{[b_I^{(-)}]}^{(n_1,n_2)} \leq 1.
\label{nondege}
\end{equation}
The reason is as follows.\footnote{
We tried to prove the non-degeneracy of $\phi_I$'s by using a
graphical consideration similar to one in the proof of
Theorem\ref{prop5}. Although for several simple cases we succeeded in
proving that the non-degeneracy is necessary for
(\ref{eq1})-(\ref{eq3}),(\ref{eq4}),(\ref{eq5}) to have a solution,
we does not have a complete proof for general cases yet.
} \\
 (i) The solution of (\ref{eq1}),(\ref{eq2})(\ref{eq3}),(\ref{eq4}),(\ref{eq5}) is
clearly included in solutions of
(\ref{eq1}),(\ref{eq2}),(\ref{eq4}),(\ref{eq5}). 
The non-degeneracy of the solutions of
(\ref{eq1}),(\ref{eq2}),(\ref{eq4}),(\ref{eq5}) is the very same
one considered in \cite{Nakajima}. See the argument at the end of section
\ref{n=2theory} and above discussions. In this case, the non-degeneracy
is certified. \\
(ii) It is clear that the degenerate solutions do not contribute to
the path integral for the partition function, because 
the factor $\prod_{I \neq J}(\phi_I - \phi_J)$ in (\ref{formula})
becomes zero if there are degenerate solutions of
$\phi_I$ \cite{Nekrasov}. \\
%

\begin{figure}
\begin{minipage}{4cm}
 \centering
 \includegraphics[width=20mm,clip]{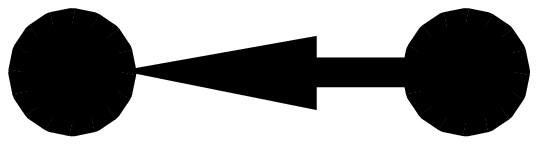}
 \caption{ $A_{z_1}$ }
 \label{fa1}
\end{minipage}
\begin{minipage}{4cm}
 \centering
 \includegraphics[width=20mm,clip]{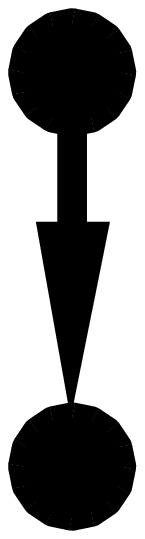}
 \caption{ $A_{z_2}$ }
 \label{fa2}
\end{minipage}
\begin{minipage}{4cm}
 \centering
 \includegraphics[width=20mm,clip]{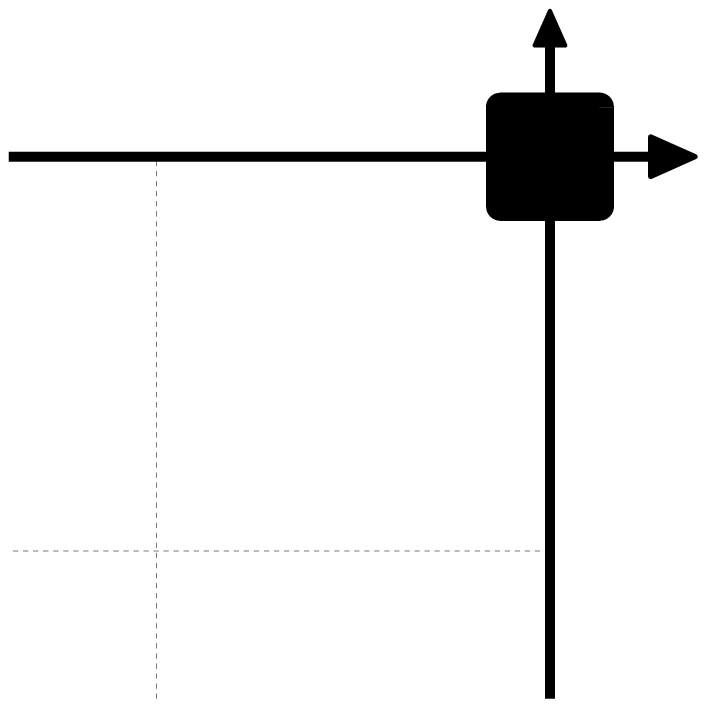}
 \caption{ $q_{\dot 1}$ }
 \label{fq}
\end{minipage}
\end{figure}

Let us give graphical interpretations of
(\ref{a1n}),(\ref{a2n}),(\ref{qn}).
\begin{flushleft}
\begin{itemize}
 \item $A_{z_1} \ {}_{\{l,(n_1,n_2)\} \{l,(n_1+1,n_2)\}}$ corresponds to
       a left-arrow connecting $(n_1,n_2)$ and
       $(n_1+1,n_2)$ in $C_{{\cal I}(l)}$. See Fig.\ref{fa1}. The number
       of non-trivial real components, ${}^{\#}\{A_{z_1}\}$, is given by two
       times of the number of the left-arrows. 
 \item $A_{z_2} \ {}_{\{l,(n_1,n_2)\} \{l,(n_1,n_2+1)\}}$ corresponds to
       a down-arrow connecting $(n_1,n_2)$ and $(n_1,n_2+1)$
       in $C_{{\cal I}(l)}$. See Fig.\ref{fa2}. The number of
       nontrivial components, ${}^{\#}\{A_{z_2}\}$, is given by two times of
       the number of the down-arrows.
 \item $q_{\dot 1} \ {}_{I=\{l,(0,0)\} J={\cal I}(l)}$ corresponds to the
       origin $(0,0)$ in $C_{{\cal I}(l)}$. See Fig.\ref{fq}. The
       number of non-trivial components, ${}^{\#}\{q\}$, is given by $2$.
\end{itemize}
\end{flushleft}

The total number of undetermined  
real variables is ${}^{\#}\{A_{z_1}\} +
{}^{\#}\{A_{z_2}\} + {}^{\#}\{q\}$.\\

\begin{figure}[hbt]
\begin{minipage}{5cm}
 \centering
 \includegraphics[width=30mm,clip]{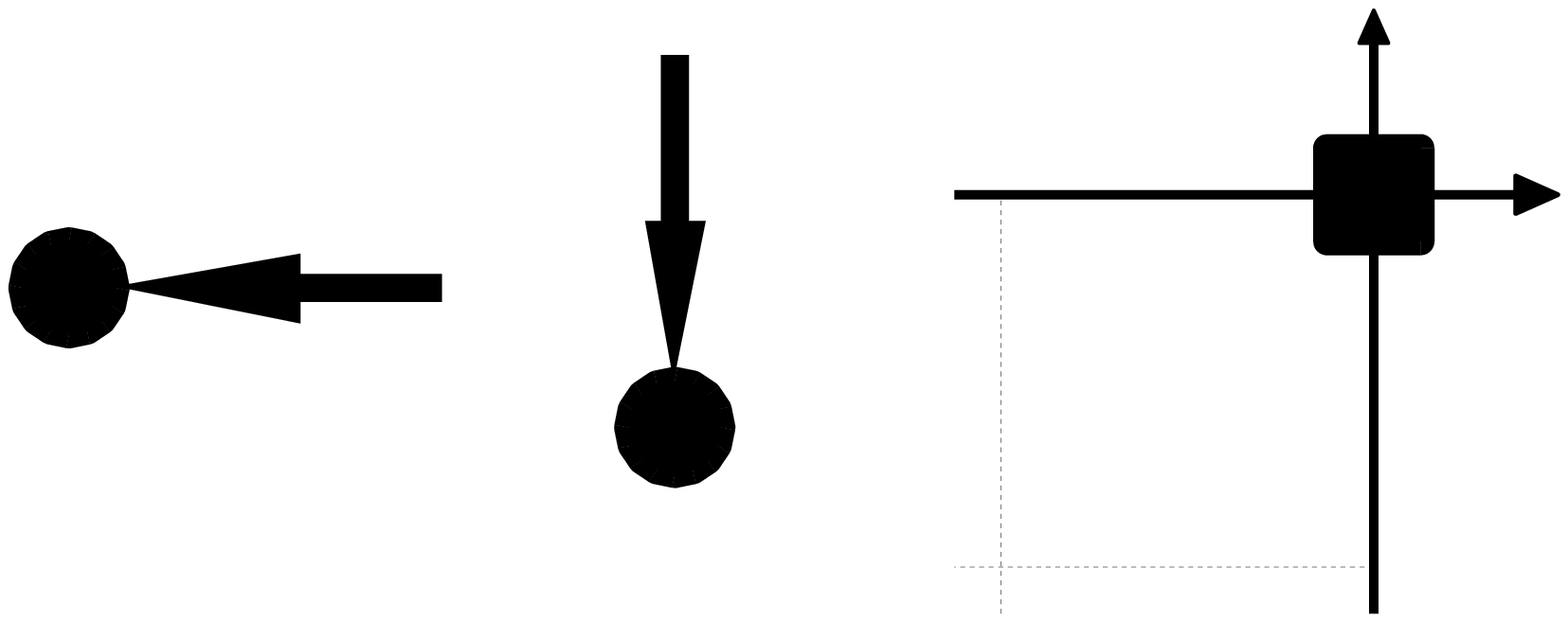}
 \caption{ Eq.(\ref{aarn}) }
 \label{faar}
\end{minipage}
\begin{minipage}{5cm}
 \centering
 \includegraphics[width=30mm,clip]{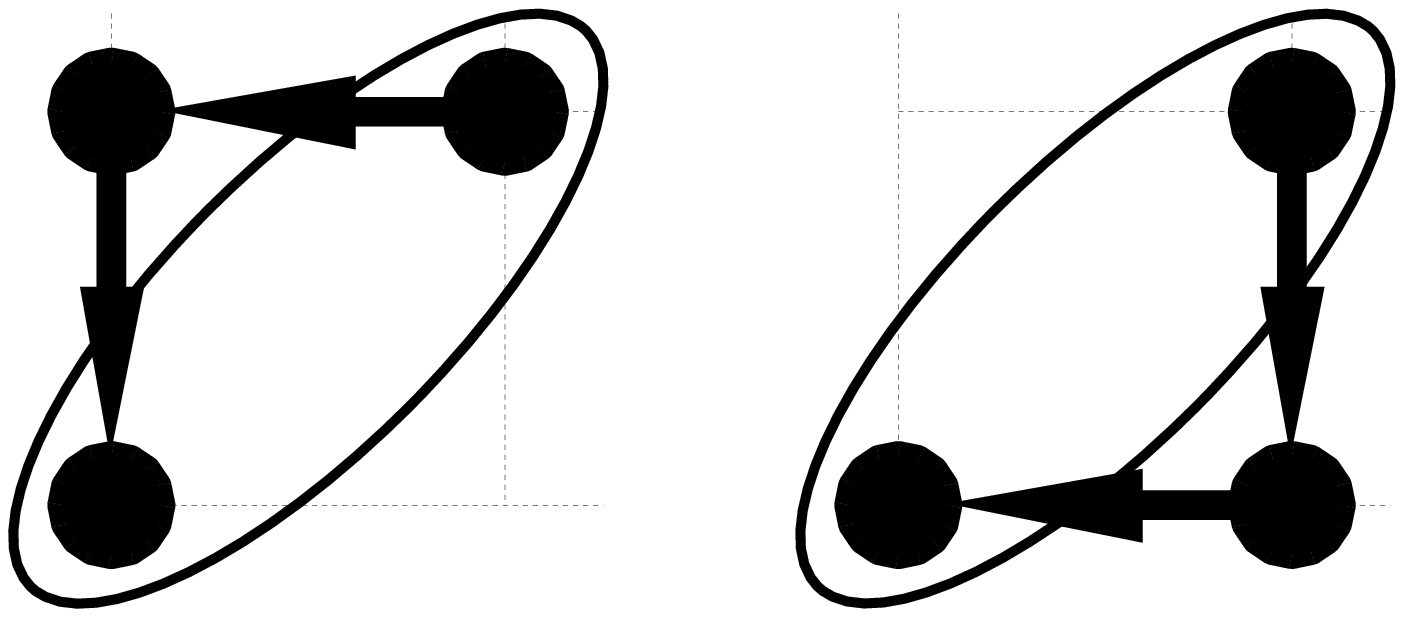}
 \caption{ Eq.(\ref{aacn}) }
 \label{faac}
\end{minipage}
\begin{minipage}{5cm}
 \centering
 \includegraphics[width=30mm,clip]{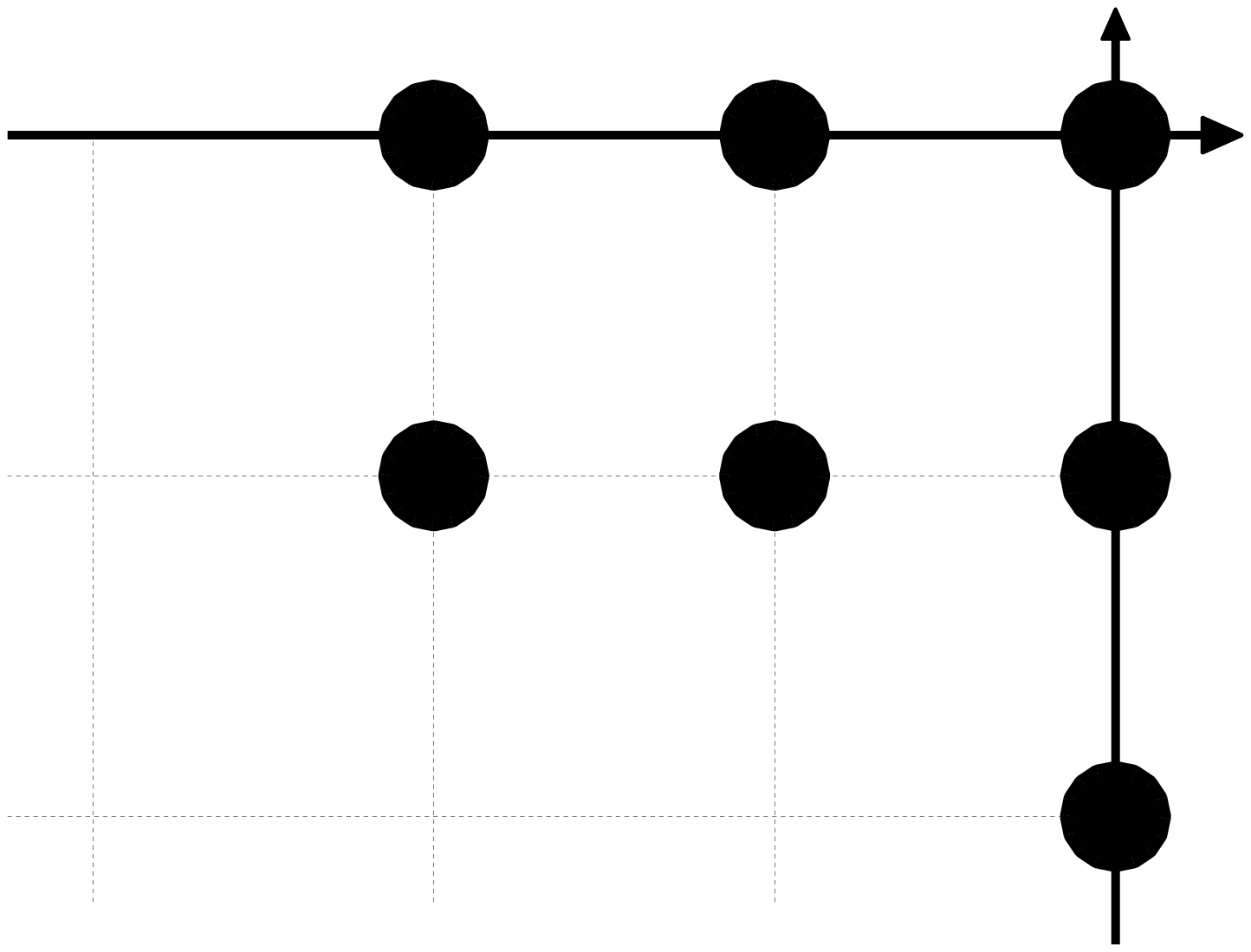}
 \caption{ $U(1)$ gauge symmetry }
 \label{fu1}
\end{minipage}
\end{figure}

Also graphical meanings of 
equations (\ref{aarn}),(\ref{aacn}) and the
residual gauge symmetry $U(1)^{N}$ are given as follows.
\begin{flushleft}
\begin{itemize}
 \item Each equation of (\ref{aarn}) corresponds to 
       ending points of left-arrow or down-arrow or the origin in
       $C_{{\cal I}(l)}$. In other words, each point $C_{{\cal I}(l)}$
       corresponds to each equation of (\ref{aarn}).
See Fig.\ref{faar}.  The number of nontrivial
       constraints, ${}^{\#}\{\mbox{Eq.(\ref{aarn})}\}$ is
       given by the number of points.
 \item Each equation of (\ref{aacn}) corresponds to a
       hook connecting $(n_1,n_2)$ and $(n_1+1,n_2+1)$, which includes
       a intermediating point $(n_1+1,n_2)$ or $(n_1,n_2+1)$, in $C_{{\cal
       I}(l)}$. See Fig.\ref{faac}. The number of nontrivial
       constraints, ${}^{\#}\{\mbox{Eq.(\ref{aacn})}\}$, is given by two times
       of the number of hooks.
 \item Each $U(1)$ factor of the residual gauge symmetry $U(1)^N$
       corresponds to each point $(n_1,n_2)$ in $C_{{\cal I}(l)}$. See
       Fig.\ref{fu1}. The number of the degrees of the residual gauge
       symmetry $U(1)^N$, denoted by ${}^{\#}\{U(1)\}$, is given by the number
       of points.
\end{itemize}
\end{flushleft}
The total number of real constraints is ${}^{\#}\{\mbox{Eq.(\ref{aarn})}\} +
{}^{\#}\{\mbox{Eq.(\ref{aacn})}\} + {}^{\#}\{U(1)\}$. \\

\begin{figure}
 \centering
 \includegraphics[width=30mm,clip]{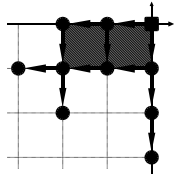}
 \caption{ A quadrangulation may include some segments which do not make faces.}
 \label{fquad}
\end{figure}

Now let us prove the next theorem. 
\begin{theorem}
Let $N$ be a finite natural number.
If and only if $C_{{\cal I}(l)}$ is a Young diagram,
(\ref{eq1}),(\ref{eq2}),(\ref{eq3}),(\ref{eq4}),(\ref{eq5}) has a solution,
and the solution is an isolated one.
\label{prop5}
\end{theorem}
(proof) \\
{}From theorem \ref{prop1}-\ref{propD}, it is enough to show that
if and only if $C_{{\cal I}(l)}$ is a Young diagram,
(\ref{aarn}) and (\ref{aacn}) has only an isolated solution.
Consider a graph $C_{{\cal I}(l)}$ as a {\it quadrangulation} of a $2$
dimensional surface.
Here we admit {\it quadrangulations} to include some segments which do
not make faces, like the graph in Fig. \ref{fquad}.\footnote{
If one considers a dual graph, then one finds that
the dual graph gives a quadrangulation of a $2$ dimensional surface
in the usual meaning.
The dual graph is obtained from the original graph by
replacing original points by dual faces and original segments connecting
original points by dual segments gluing
dual faces.}
We start with cases, where $2$ dimensional surfaces have no hole.
Recall the well-known formula for the Euler number $\chi$ of graphs,
\begin{equation}
\chi = 2 - 2h -b = {}^{\#} \{ \mbox{points} \} - {}^{\#} \{
\mbox{segments} \} + {}^{\#} \{ \mbox{faces} \},
\label{euler}
\end{equation}
where $h$ denotes the number of handles of graphs,
and $b$ denotes the number of boundaries of graphs.

\begin{figure}
\begin{minipage}{70mm}
 \centering
 \includegraphics[width=40mm,clip]{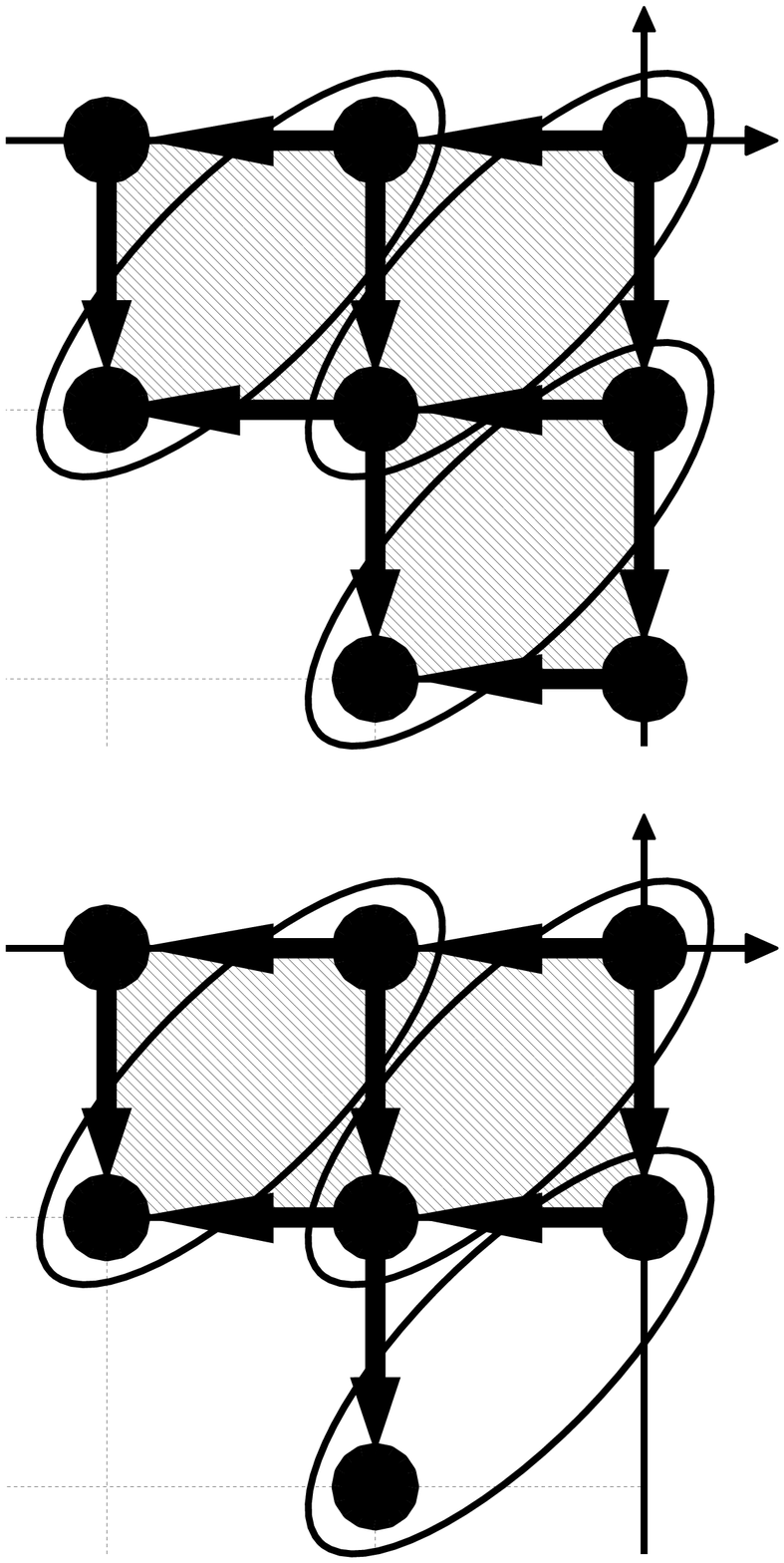}
 \caption{ Young diagram and \\ \hspace{19mm} variant diagram} 
 \label{fgy}
\end{minipage}
\hspace{5mm}
\begin{minipage}{70mm}
 \centering
 \includegraphics[width=40mm,clip]{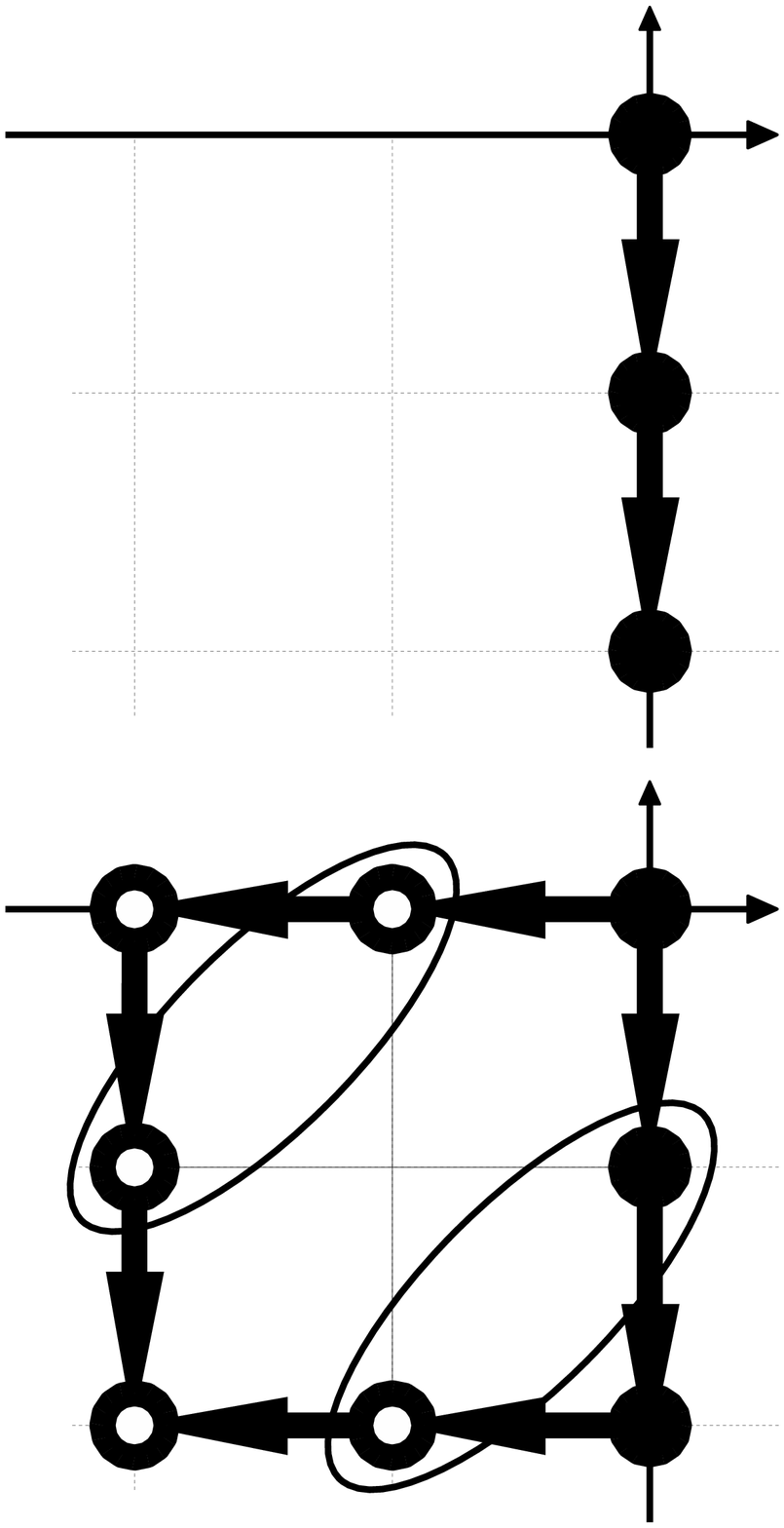}
 \caption{ Graph without a hole and \\ \hspace{19mm} graph with a hole }
 \label{fgnsc}
\end{minipage}
\end{figure}

In our case, $h = 0$ and $b = 1$.
Then we obtain,
\begin{equation}
\chi = 1 = {}^{\#} \{ \mbox{points} \} - {}^{\#} \{ \mbox{segments} \} + {}^{\#} \{ \mbox{faces} \}.
\end{equation}
Notice that
\begin{equation}
{}^{\#} \{ \mbox{points} \} = {}^{\#}\{\mbox{Eq.(\ref{aarn})}\} = {}^{\#}\{U(1)\},
\label{points}
\end{equation}
and
\begin{equation}
{}^{\#} \{ \mbox{segments} \} = \frac{{}^{\#}\{A_{z_1}\} + {}^{\#}\{A_{z_2}\}}{2}.
\label{edges}
\end{equation}
Also one sees that
\begin{equation}
{}^{\#} \{ \mbox{faces} \} \leq \frac{{}^{\#}\{\mbox{Eq.(\ref{aacn})}\}}{2},
\label{faces}
\end{equation}
and that, in (\ref{faces}), the equation holds when the graph $C_{{\cal I}(l)}$ is a Young diagram. See Fig.\ref{fgy}.
Then we obtain
\begin{eqnarray}
 & & \left( {}^{\#}\{A_{z_1}\} +
{}^{\#}\{A_{z_2}\} + {}^{\#}\{q\} \right) - \left( {}^{\#}\{\mbox{Eq.(\ref{aarn})}\} +
{}^{\#}\{\mbox{Eq.(\ref{aacn})}\} + {}^{\#}\{U(1)\} \right) \nonumber \\
 &=& 2 {}^{\#} \{ \mbox{segments} \} + 2 - 2 {}^{\#} \{ \mbox{points} \} - {}^{\#}\{\mbox{Eq.(\ref{aacn})}\} \nonumber \\
 &\leq& - 2 \left( {}^{\#} \{ \mbox{points} \} - {}^{\#} \{ \mbox{segments} \} + {}^{\#} \{ \mbox{faces} \} \right) + 2 \nonumber \\
 &=& - 2 + 2 \nonumber \\
 &=& 0.
\end{eqnarray}
{}From this, we find that if and only if $C_{{\cal I}(l)}$ is
a Young diagram, we can have a solution to (\ref{aarn}),(\ref{aacn}),
and that the solution is an isolated one.

Now let us turn to a case, where
$C_{{\cal I}(l)}$ has some holes.
A diagrams with holes is constructed from one without holes
by adding pieces of diagrams.
For example, see Fig.\ref{fgnsc}.
In Fig.\ref{fgnsc}, some white dots are added to make a hole.
Under this operation, the number of undetermined variables increases by
\begin{equation}
\Delta \ {}^{\#} \{ \mbox{undetermined variables} \} = \Delta \
 {}^{\#}\{A_{z_1}\} + 
 \Delta \ {}^{\#}\{A_{z_2}\} = 2 \times 4 + 2 \times 2  = 12.
\end{equation}
On the other hand, the number of constraints increases by
\begin{equation}
\Delta \ {}^{\#} \{ \mbox{constraints} \} = \Delta \
 {}^{\#}\{\mbox{Eq.(\ref{aarn})}\} + 
 \Delta \ {}^{\#}\{\mbox{Eq.(\ref{aacn})}\} + \Delta \ {}^{\#}\{U(1)\} = 5 + 2 \times 2 + 5 = 14.
\end{equation}
As implied by the above example,
one can show that ``puncture'' operations make the number of constraints greater
 than that of undetermined variables in general.
We conclude that if $C_{{\cal I}(l)}$ has some holes,
then (\ref{aarn}),(\ref{aacn}) have no solution.

We have finished the proof for Theorem\ref{prop5}. $\blacksquare$\\

As mentioned in the top of this section,
we have shown that
(\ref{eq1}),(\ref{eq2}),(\ref{eq3}),(\ref{eq4}),(\ref{eq5}) have only
isolated solutions, and the solutions are expressed by the Young
diagrams. 

At the end of this section,
we comment on the case that the $q$ are not square matrices. 
Let us compare above cases with
the case of ${\mathbb C}^{[n]}$ and the ADHM data
for usual U(N) instanton.
We have investigated the case that $q_{\dot{\alpha}}$ and 
$q_{\dot{\alpha}}^{\dagger}$ are
$N \times N$ square matrices.
It is clear that the above theorem is valid even if $q_{\dot{\alpha}}$
and  $q_{\dot{\alpha}}^{\dagger}$ are $N \times k$ and $k \times N$
for arbitrary  $k \in {\mathbb Z}$,
respectively.
In this case, our equations (\ref{eq1}) - (\ref{eq2}) are 
ADHM equations corresponding to U(N) instanton of $k$ instanton number
with Dirac equation reduced to $0$ dimension.
The Dirac equation (\ref{eq3}) makes no nontrivial
equations when we introduce $\zeta$.
Then, our models are completely equivalent to the case of ADHM
equations with fixed point equations of torus action,
that is discussed in Nakajima's lecture note \cite{Nakajima}
and others \cite{Nekrasov,Nakajima-Yoshioka,Nekrasov-Okounkov}.
The proof for the correspondence with ADHM data and the Young diagrams
is given by \cite{Nakajima}.
In this light, our proof in this section is a new version for the 
Nakajima's theorem.
We solved the fixed point equation of the torus action directly.
By virtue of the concrete solution, the correspondence
between fields components, ADHM equations and Young diagrams are clarified.

%
%
\section{Localization Theorem} \label{localization}
Though, in this paper, we does not perform the summation of the solutions nor
obtain the partition function of our model,
we make comment on the localization theorem 
\cite{Lo-Ne-Sh}-\cite{AB}, which is a
powerful tool for the calculation of path integral of cohomological
field theories,
in order to explain our motivation.
To carry out the calculation of infinite $N$ case, that is N.C.${\mathbb R}^4$
case, is difficult.
Therefore we consider the toy model that is given by the same type
Lagrangian of section \ref{n=2theory} but its all fields are finite
$N\times N$ matrices.

For our purpose,
one of the most suitable expression of the localization theorem
is one given in \cite{Bruzzo,Bruzzo1}.
This is expressed as follows.

Let
${\tilde \delta}$ be the deformed BRS transformation defined in section \ref{equation}.
As explained in section \ref{n=2theory},
the action $S$ is given by a BRS exact function.
Now we redefine the action as
\begin{equation}
S = {\tilde \delta} \Psi(\phi,{\cal B},{\cal F}).
\end{equation}
The difference between ${\hat \delta} \Psi$ and ${\tilde \delta} \Psi$
 causes no effect to the path integral, because the integral of
 equivariant cohomology is equal to that of original cohomology.
Here we have used the notation ${\cal
 B},{\cal F}$ to denote the BRS doublet fields collectively.
Then the localization theorem tells us that
\begin{equation}
Z = \int \frac{D \phi}{U(N)} D {\cal B} D {\cal F} e^{-{\tilde \delta} \Psi} = \int
 \prod_{I=1}^{N} d \phi_I \frac{\prod_{I \neq J}(\phi_I -
 \phi_J)}{Sdet^{\frac{1}{2}} {\cal L}}.
\label{formula}
\end{equation}
$\phi_I$ are the eigenvalues of $\phi$, and
the superdeterminant $Sdet {\cal L}$ is defined by
\begin{equation}
Sdet {\cal L} = Sdet \left(
\begin{array}{cc}
\frac{\partial (Q)_{\cal B}}{\partial {\cal F}} & 
\frac{\partial (Q)_{\cal B}}{\partial {\cal B}} \\
\frac{\partial (Q)_{\cal F}}{\partial {\cal F}} &
\frac{\partial (Q)_{\cal F}}{\partial {\cal B}}
\end{array}
\right),
\end{equation}
where $(Q)_{\cal B}$ and $(Q)_{\cal F}$ are defined by
the representation of the deformed BRS transformation ${\tilde \delta}$ on the fields ${\cal
B},{\cal F}$,
\begin{equation}
Q = (Q)_{\cal B} \frac{\partial}{\partial {\cal B}} + (Q)_{\cal F} \frac{\partial}{\partial {\cal F}}.
\end{equation}
Note that this expression is analogue of
\begin{equation}
{\tilde d} = d + i_{X},
\end{equation}
where $X$ is a vector defining the Lie derivative ${\cal L}_X$
associated with ${\cal G} \otimes T^{N+2}$ action. See
(\ref{brsA}),(\ref{brsq1}),(\ref{brsq2}).
In our case, we obtain
\begin{eqnarray}
Z &=& \int \prod_{I=1}^N d \phi_I \prod_{I \neq J} (\phi_I - \phi_J)
 \prod_{I=1}^{N} \frac{(\epsilon_1 + \epsilon_2)\{ -(\phi_I - b_I)^2 +
 \epsilon_-^2 \}}{\epsilon_1 \epsilon_2 \{ -(\phi_I - b_I)^2 +
 \epsilon_+^2 \}} \nonumber \\
 & & \prod_{I \neq J} \frac{\{ (\phi_I - \phi_J)^2 - 4 \epsilon_+^2 \}^{\frac{1}{2}}
  \{ -(\phi_I - b_J)^2 + \epsilon_-^2 \}}{\{ -(\phi_I - b_J)^2 +
  \epsilon_+^2 \} \{(\phi_I - \phi_J)^2 - \epsilon_1^2\}^{\frac{1}{2}} \{(\phi_I -
  \phi_J)^2 - \epsilon_2^2\}^{\frac{1}{2}}}, 
\label{formula'}
\end{eqnarray}
where $\epsilon_- = (\epsilon_1 - \epsilon_2) / 2$.

Some comments might be necessary.
This formula is derived by using a some version of localization theorem,
which reduces the integral $\int D {\cal B} D {\cal F}$,
and this is valid only if the BPS equations of the action
(\ref{eq1}),(\ref{eq2}),(\ref{eq3}) and
the fixed point equations of the deformed BRS symmetry
(\ref{eq4}),(\ref{eq5}) have isolated
solutions for a given value of $\phi_I$'s.
%
The integral $\int \prod_I d \phi_I$
is remained, and this should be understood as the contour
integral. In order to define an appropriate contour, 
we use $\epsilon_i \rightarrow \epsilon_i + i 0$ prescription.
The poles correspond to the
isolated solutions \cite{Lo-Ne-Sh}-\cite{Mo-Ne-Sh1}.

%
%
\section{Conclusion} \label{conclusion}
The solutions of the Seiberg-Witten monopole equations reduced to $0$ dimension
which also satisfy the fixed point equations of torus actions
were classified,
where the torus action is induced from the global symmetries.
More concretely speaking, we deformed the BRS transformation 
of the topological twisted ${\cal N}=2$ gauge theory 
on ${\mathbb R}^4$ with a hypermultiplet
to the T-equivariant derivative
by using the global symmetries.
The global symmetries contain torus actions.
Using these symmetries, the deformed BRS transformation was defined 
to satisfy the nilpotency up to the Lie derivative of the group actions.
Then we classified the solutions of the fixed point equations of these
deformed BRS transformations.

We showed that the Seiberg-Witten monopole equations are
reduced to the ADHM equations with the Dirac equation reduced to $0$
dimension
at the large N.C. parameter limit.
These equations are described by using infinite dimensional matrices.
We showed that the Dirac equation reduced to $0$ dimension is trivial
when the ADHM equations and the fixed point
equations are satisfied. 
It is known that the solutions of the ADHM equations with 
the fixed point equations 
are isolated ones,
and are classified by the Young diagrams, when matrix size is finite.
We gave a new proof of this statement, too.
Then, we found that we can perform the path integral 
by using the localization formula,
in order to get
the partition functions of the finite dimensional matrix model.
This finite dimensional matrix model is given as 
reduced theory to $0$ dimension from 
the topological twisted ${\cal N}=2$ non-Abelian gauge theory 
on ${\mathbb R}^4$ with a hypermultiplet,
because the size of matrix is truncated to finite dimension from
infinite dimension.
We gave the result of the partition function of this toy model.
The complete calculation of the partition function for 
the ${\cal N}=2$ $U(1)$ gauge theory 
on N.C. ${\mathbb R}^4$ is remained.
This calculation might reveal the relation between the Seiberg-Witten
monopole and the instanton.
We hope to report on this task elsewhere.

\appendix

%
%
\section{Convention} \label{conv}

\subsection{Complex coordinate}
We define the complex coordinate $z^i,{\bar z}^i \ (i=1,2)$ as
\begin{eqnarray}
& & z^1 = \frac{1}{\sqrt{2}}(x^1 + i x^2) \ , \ {\bar z}^1 =
 \frac{1}{\sqrt{2}}(x^1 - i x^2) \ \ , \nonumber \\
& & z^2 = \frac{1}{\sqrt{2}}(x^3 + i x^4) \ , \ {\bar z}^2 =
 \frac{1}{\sqrt{2}}(x^3 - i x^4) \ \ .
\end{eqnarray}
Also, $\partial_{z^i},\partial_{{\bar z}^i}$ are given by
\begin{eqnarray}
& & \partial_{z^1} = \frac{1}{\sqrt{2}}(\partial_1 - i \partial_2) \ , \
 \partial_{{\bar z}^1} =
 \frac{1}{\sqrt{2}}(\partial_1 + i \partial_2) \ \ , \nonumber \\
& & \partial_{z^2} = \frac{1}{\sqrt{2}}(\partial_3 - i \partial_4) \ , \
 \partial_{{\bar z}^2} =
 \frac{1}{\sqrt{2}}(\partial_3 + i \partial_4) \ \ .
\end{eqnarray}
Then, we obtain
\begin{equation}
\partial_{z^i} z^j = \delta_i^j \ , \ \partial_{{\bar z}^i} {\bar z}^j =
 \delta_i^j \ \ .
\end{equation}

\subsection{Spinor index}
$\epsilon^{\alpha \beta}$,$\epsilon^{{\dot \alpha} {\dot \beta}}$ and $\epsilon_{\alpha \beta}$,$\epsilon_{{\dot \alpha} {\dot \beta}}$ are defined by
\begin{equation}
\epsilon^{\alpha \beta} = \epsilon^{{\dot \alpha} {\dot \beta}} = \left(
\begin{array}{cc}
0 & +1 \\
-1 & 0
\end{array}
\right) \ , \ 
\epsilon_{\alpha \beta} = \epsilon_{{\dot \alpha} {\dot \beta}} = \left(
\begin{array}{cc}
0 & -1 \\
+1 & 0
\end{array}
\right) \ \ .
\end{equation}
In other words, $\epsilon_{\alpha \beta}$,$\epsilon_{{\dot \alpha} {\dot \beta}}$ are defined to be the inverses
of $\epsilon^{\alpha \beta}$,$\epsilon^{{\dot \alpha} {\dot \beta}}$,
\begin{equation}
\epsilon^{\alpha \beta} \epsilon_{\beta \gamma} =
 \delta^\alpha_\gamma \ , \ \epsilon^{{\dot \alpha} {\dot \beta}}
 \epsilon_{{\dot \beta} {\dot \gamma}} =
 \delta^{\dot \alpha}_{\dot \gamma} \ \ . 
\end{equation}
Then a spinor with upper indices and a spinor with lower indices are
related as,
\begin{eqnarray}
& & \psi^\alpha = \epsilon^{\alpha \beta} \psi_\beta \ , \ \psi_\alpha =
 \epsilon_{\alpha \beta} \psi^\beta \ \ , \nonumber \\
& & \psi^{\dot \alpha} = \epsilon^{{\dot \alpha} {\dot \beta}}
 \psi_{\dot \beta} \ , \ \psi_{\dot \alpha} =
 \epsilon_{{\dot \alpha} {\dot \beta}} \psi^{\dot \beta} \ \ .
\end{eqnarray}

We use the following definition for the $4$ dimensional Pauli
matrix $\sigma^{\mu} \ (\mu=1,2,3,4)$,
\begin{eqnarray}
& & (\sigma^{\mu})_{\alpha {\dot{\alpha}}} 
= \left( \sigma^1 , \sigma^2 , \sigma^3 , \sigma^4 \right) 
= \left( i {\bf 1} \ , \ - {\vec \tau} \right) \ ,
\nonumber \\
& & ({\bar \sigma}^{\mu})^{{\dot \alpha} \alpha} 
= \left( {\bar \sigma}^1 , {\bar \sigma}^2 , {\bar \sigma}^3 , {\bar \sigma}^4 \right) 
= \left( i {\bf 1} \ , \ + {\vec \tau} \right) \ , 
\end{eqnarray}
where
\begin{equation}
{\vec \tau} = \left(
\left( 
\begin{array}{cc}
0 & +1 \\
+1 & 0       
\end{array}
\right)
 , 
\left( 
\begin{array}{cc}
0 & -i \\
+i & 0       
\end{array}
\right)
 , 
\left( 
\begin{array}{cc}
+1 & 0 \\
0 & -1       
\end{array}
\right)
\right) \ .
\end{equation}
We define $\sigma^{\mu\nu} , {\bar \sigma}^{\mu\nu}$ as
\begin{eqnarray}
& & {(\sigma^{\mu\nu})_\alpha}^\beta 
= \frac{i}{4} {\left(
\sigma^\mu {\bar \sigma}^\nu - \sigma^\nu {\bar \sigma}^\mu
\right)_\alpha}^\beta \ ,
\nonumber \\
& & {({\bar \sigma}^{\mu\nu})^{\dot \alpha}}_{\dot \beta} 
= \frac{i}{4} {\left(
{\bar \sigma}^\mu \sigma^\nu - {\bar \sigma}^\nu \sigma^\mu 
\right)^{\dot \alpha}}_{\dot \beta} \ .
\end{eqnarray}
{}From this definition, $\sigma^{\mu\nu}$ and ${\bar \sigma}^{\mu\nu}$
satisfy the anti selfdual relation and the selfdual relation respectively,
\begin{equation}
\sigma^{\mu\nu} = - * \sigma^{\mu\nu} \ , \ {\bar \sigma}^{\mu\nu}
 = + * {\bar \sigma}^{\mu\nu} \ \ .
\end{equation}

\subsection{$\dagger$ symbol}
For a scalar matrix $M$ and a vector matrix $M_{\mu}$,
the symbol $\dagger$ denotes the usual hermite conjugation for them,
\begin{equation}
M^{\dagger} = M^{* T} \ , \ {M_\mu}^{\dagger} = {M_\mu}^{* T},
\end{equation}
where the symbol $*$ denotes the complex conjugation and the symbol $T$
denotes the transposition. On the other hand, for an undotted spinor
matrix $M_\alpha$ and a dotted spinor matrix $M_{\dot \alpha}$, 
${M_\alpha}^\dagger$ and ${M_{\dot \alpha}}^\dagger$ are defined by,
\begin{equation}
{M_\alpha}^\dagger = \epsilon^{\alpha \beta} {M_\beta}^{* T} \ , \
 {M_{\dot \alpha}}^\dagger = \epsilon^{{\dot \alpha} {\dot \beta}}
 {M_{\dot \beta}}^{* T} \ \ .
\end{equation}
This definition makes ${M_\alpha}^\dagger$ and ${M_{\dot
\alpha}}^\dagger$ to transform in the same rules as $M_\alpha$ and
$M_{\dot \alpha}$ under $SU(2)_L$ and $SU(2)_{R (R')}$ respectively.


%
%

\end{document}